\documentclass[prx,amsmath,amssymb, notitlepage, twocolumn,floatfix,nofootinbib,superscriptaddress, longbibliography]{revtex4-1}
\usepackage{amsmath}
\usepackage{mathtools} 
\usepackage{tabularx,graphicx}
\usepackage{epstopdf}
\usepackage{graphicx}
\usepackage{enumerate}
\usepackage{spectralsequences} 
\usepackage{latexsym}
\usepackage{amssymb}
\usepackage{tikz-cd}
\usepackage{amsmath}
\usepackage{color, colortbl}
\usepackage{psfrag}
\usepackage{bbm}
\usepackage{bm}
\usepackage{titlesec}
\usepackage{dsfont}
\usepackage{feynmp}
\usepackage{slashed}
\usepackage{multirow}
\usepackage{framed}
\usepackage{braket}
\newcommand{\mb}[1]{\mathbf{#1}}

\newcommand{\beq}{\begin{eqnarray}}
\newcommand{\eeq}{\end{eqnarray}}

\newcommand{\Z}{\mathbb{Z}}
\newcommand{\U}{\mathrm{U}(1)}
\newcommand{\bs}{\boldsymbol}

\DeclareMathOperator{\Tr}{Tr}
\DeclareMathOperator{\tr}{tr}

\newcommand{\bsp}{\begin{split}}
\newcommand{\esp}{\end{split}}
\newcommand{\const}{{\rm const}}

\newcommand{\ie}{{i.e., }}

\newcommand{\nn}{\nonumber \\}
\newcommand{\R}{\mathcal{R}}
\newcommand{\E}{\mathcal{E}}
\newcommand{\HH}{\mathcal{H}}
\newcommand{\one}{\mathbbm{1}}
\newcommand{\ketbra}[2]{\ket{#1}\!\bra{#2}}
\usepackage{color}
\definecolor{darkblue}{rgb}{0.,0.,0.4}
\definecolor{darkred}{rgb}{0.5,0.,0.}
\definecolor{BlueViolet}{RGB}{138,43,226}
\definecolor{SkyBlue}{RGB}{30,144,255}
\definecolor{DarkGreen}{RGB}{0,100,0}
\usepackage[pdftex,colorlinks=true,linkcolor=darkblue,citecolor=blue,urlcolor=darkred]{hyperref}
\usepackage[normalem]{ulem}

\makeatletter
\newsavebox{\@brx}
\newcommand{\llangle}[1][]{\savebox{\@brx}{\(\m@th{#1\langle}\)}%
  \mathopen{\copy\@brx\kern-0.5\wd\@brx\usebox{\@brx}}}
\newcommand{\rrangle}[1][]{\savebox{\@brx}{\(\m@th{#1\rangle}\)}%
  \mathclose{\copy\@brx\kern-0.5\wd\@brx\usebox{\@brx}}}
\makeatother
\usepackage{amsthm}
\theoremstyle{plain}
\newtheorem*{theorem*}{Theorem}
\newtheorem{theorem}{Theorem}
\newtheorem{corollary}{Corollary}[theorem]
\newtheorem{conjecture}{Conjecture}
\newtheorem{definition}{Definition}

\newcommand{\add}[1]{ {    #1 }}

\newcommand{\chong}[1]{ { \color{darkred} \footnotesize (\textsf{CW}) \textsf{\textsl{#1}} }}

\begin{document}
\title{Strong-to-Weak Spontaneous Symmetry Breaking in Mixed Quantum States}
	
\author{Leonardo A. Lessa}
\thanks{These authors contributed equally to this work.}
\affiliation{Perimeter Institute for Theoretical Physics, Waterloo, Ontario N2L 2Y5, Canada}
\affiliation{Department of Physics and Astronomy, University of Waterloo, Waterloo, Ontario N2L 3G1, Canada}

\author{Ruochen Ma}
\thanks{These authors contributed equally to this work.}
\affiliation{Kadanoff Center for Theoretical Physics, University of Chicago, Chicago, Illinois 60637, USA}
\author{Jian-Hao Zhang}
\thanks{These authors contributed equally to this work.}
\affiliation{Department of Physics, The Pennsylvania State University, University Park, Pennsylvania 16802, USA}
\affiliation{Department of Physics and Center for Theory of Quantum Matter, University of Colorado, Boulder, Colorado 80309, USA}
\author{Zhen Bi}
\affiliation{Department of Physics, The Pennsylvania State University, University Park, Pennsylvania 16802, USA}
\author{Meng Cheng}
\affiliation{Department of Physics, Yale University, New Haven, Connecticut  06511-8499, USA}
\author{Chong Wang}
\affiliation{Perimeter Institute for Theoretical Physics, Waterloo, Ontario N2L 2Y5, Canada}

\begin{abstract}
Symmetry in mixed quantum states can manifest in two distinct forms: \textit{strong symmetry}, where each individual pure state in the quantum ensemble is symmetric with the same charge, and \textit{weak symmetry}, which applies only to the entire ensemble. This paper explores a novel type of spontaneous symmetry breaking (SSB) where a strong symmetry is broken to a weak one. While the SSB of a weak symmetry is measured by the long-ranged two-point correlation function, the strong-to-weak SSB (SW-SSB) is measured by the \textit{fidelity correlator}. We prove that SW-SSB is a universal property of mixed-state quantum phases, in the sense that the phenomenon of SW-SSB is robust against symmetric low-depth local quantum channels. We also show that the symmetry breaking is ``spontaneous'' in the sense that the effect of a local symmetry-breaking measurement cannot be recovered locally. We argue that a thermal state at a nonzero temperature in the canonical ensemble (with fixed symmetry charge) should have spontaneously broken strong symmetry. Additionally, we study non-thermal scenarios where decoherence induces SW-SSB, leading to phase transitions described by classical statistical models with bond randomness. In particular, the SW-SSB transition of a decohered Ising model can be viewed as the ``ungauged'' version of the celebrated toric code decodability transition. We confirm that, in the decohered Ising model, the SW-SSB transition defined by the fidelity correlator is the only physical transition in terms of channel recoverability. We also comment on other (inequivalent) definitions of SW-SSB, through correlation functions with higher R\'enyi indices.
\end{abstract}

\maketitle

\tableofcontents

\section{Introduction}

The notion of \textit{spontaneous symmetry breaking} (SSB) is a cornerstone of modern physics \cite{LandauLifshitz, McGreevy2022}. SSB in quantum physics has been well understood for pure states -- typically the ground states of some many-body Hamiltonians -- and for Gibbs thermal states $\rho=e^{-\beta H}/Z$. Surprisingly, recent advances in understanding phases of matter in mixed quantum states have revealed a novel type of SSB dubbed strong-to-weak SSB (SW-SSB) \cite{Lee_2023, Ma:2023rji}. This work aims to establish some key universal aspects of SW-SSB and associated phase transitions.

\begin{figure}
    \centering
    \includegraphics[width=.48\textwidth]{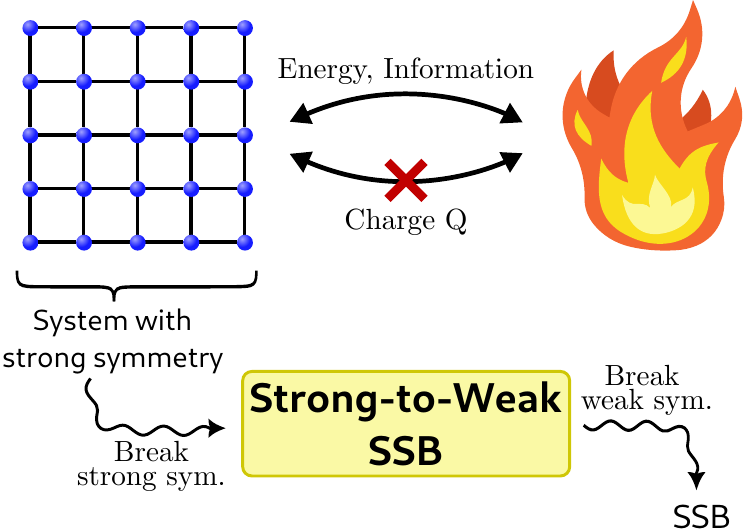}
    \caption{Main scenario considered in this work: a mixed state system $\rho$ with strong symmetry $U \rho \propto \rho$, such as one interacting with a (thermal) environment without exchanging charge, that spontaneously breaks the strong symmetry but not the weak symmetry. When this occurs, we call it a strong-to-weak SSB state, detected via a long-range correlator $F_O(x, y) \sim O(1)$. If the weak symmetry is further broken down, then the system exhibits the usual SSB, $\langle O(x) O^\dag(y)\rangle_\rho \sim O(1)$.}
    \label{fig:MainFigure}
\end{figure}

In quantum mechanics, symmetry acts on pure states via multiplication by the corresponding unitary (or anti-unitary) $U$ and acts on operators via conjugation. For an open quantum system described by a density operator $\rho$, however, depending on the coupling to the environment there are two ways that symmetry can act~\cite{BucaProsen2012, AlbertJiang2014, Albert2018, Lieu2020}. One can have a \textit{strong symmetry} (also called \textit{exact symmetry}), defined as
\begin{align}
U\rho=e^{i\theta}\rho,
\end{align}
which means that, when $\rho$ is viewed as an ensemble, every member state carries the same charge under the symmetry. Alternatively, the symmetry can be just \textit{weak} (or \textit{average}), and not strong, defined as
\begin{align}
U\rho\ne e^{i\theta}\rho,~~U\rho U^\dag=\rho,
\end{align}
which can happen in an ensemble whose member states have well-defined but different symmetry charges. Physically, strong symmetry arises when the system does not exchange charges with the environment, {as illustrated in Fig. \ref{fig:MainFigure}}. These concepts are deeply rooted in equilibrium statistical mechanics: canonical ensembles have ``strong'' particle number conservation symmetry, while grand canonical ensembles are only weakly symmetric.

Given these two ways that symmetries can be realized in a mixed state, there is a need to revisit the notion of spontaneous symmetry breaking, which serves as a fundamental organizing principle in quantum many-body physics. In order to clarify the matter, consider a many-body state $\rho$ with global symmetry $G$, and a charged local operator $O(x)$. Conventional SSB can be defined through the long-range order $\Tr[ O(x)O^\dag(y)\rho]\rightarrow \text{const} \neq 0$ when $|x-y|\rightarrow \infty$. In this case, regardless of whether the symmetry action $g$ is strong or weak, it is {spontaneously} broken to nothing. 
Now given the necessity to distinguish strong and weak symmetries,  
there is a distinct possibility that a strong symmetry can be spontaneously broken into a weak one. This phenomenon of strong-to-weak SSB (SW-SSB) is only possible for mixed states. {The relation between the two notions of SSB is illustrated in Fig. \ref{fig:MainFigure}.} In Sec.~\ref{Sec: def} we carefully define the notion of SW-SSB and study its key properties. Our central result, dubbed the \textit{Stability Theorem} (Thms.~\ref{thm:SW-SSB_stability} and \ref{thm:SW-SSB_stability2}), establishes SW-SSB as a universal property of mixed-state quantum phases that is robust against low-depth symmetric local quantum channels.

While it might sound unfamiliar, recall for spin glass order in disordered systems, the ensemble-averaged order parameter vanishes $\overline{\langle O(x)\rangle}=0$, but each individual state in the disordered ensemble has symmetry-breaking order, reflected in the Edward-Anderson \cite{EdwardAnderson} order parameter $\overline{|\langle O(x)\rangle|}\neq0$. In our language, this means that the global symmetry (under which $O(x)$ is charged) is broken as an exact (strong) symmetry, but is preserved as an average (weak) symmetry. We will explain in Sec.~\ref{Sec:fidelity} that our SW-SSB is, in a precise sense, a mixed-state generalization of the Edward-Anderson type of spin glass order.

Another familiar example comes from equilibrium quantum statistical mechanics: the fact that canonical and grand canonical ensembles are completely equivalent in the thermodynamic limit in fact is nothing but SW-SSB (we will explain this point in Sec.~\ref{sec:thermal}). Therefore, to some extent, SW-SSB should be viewed as a fundamental property of thermal states with strong symmetry. 

Recently, non-thermal mixed states have received much attention thanks to the rapid progress in engineering and controlling complex many-body states in quantum devices, motivating many investigations into quantum phases in mixed states \cite{Lee_2023, bao2023mixedstate, fan2024diagnostics, Zou_2023, chen2023symmetryenforced, chen2024separability, sang2023mixedstate, sang2024stability, rakovszky2024defining, Lu_2023, lee2022decoding, Zhu_2023, lessa2024mixedstate, chen2024unconventional, wang2023intrinsic, wang2024anomaly, sohal2024noisy, Kawabata_2024, zhou2023reviving, hsin2023anomalies, guo2024designing, lu2024disentangling}. While the full landscape of mixed state phases is still largely uncharted, novel topological states protected by both strong and weak symmetries have been identified, including examples that have no counterpart in pure states.
Given the central role SSB has played in the theory of the ground state (or equilibrium) phases, understanding SW-SSB is clearly a foundational issue for mixed-state quantum phases. For example, to define mixed-state symmetry-protected topological phases~\cite{deGroot2022, MaWangASPT, ZhangQiBi2022, LeeYouXu2022, Ma:2023rji, Zhang_2023, Ma:2024kma, guo2024locally, Xue:2024bkt, chirame2024stable}, it is necessary to impose the absence of any symmetry breaking, including the SW-SSB~\cite{Lee_2023, Ma:2023rji, Ma:2024kma}.

The simplest setting of SW-SSB in a non-thermal state is to start from a symmetric pure state and introduce local symmetric decoherence. In Secs.~\ref{sec:Ising} and \ref{sec:rotor} we study two such examples, one with Ising $\Z_2$ symmetry, the other with $\U$. In both cases, we find phase transitions into SW-SSB at some finite decoherence strength $p_c$, with the universality classes described by some classical statistical mechanical models with bond randomness. In particular, the celebrated decodability transition of $\Z_2$ toric code can be viewed as the ``gauged'' version (through Kramers-Wannier duality) of the Ising SW-SSB transition. For the decohered Ising model, we also show (Sec.~\ref{sec:recover}) that the only physical transition is the one involving SW-SSB. In particular, using the recently proposed local Petz recovery channel \cite{sang2024stability}, we show that two states in the same phase (both with or without SW-SSB) are two-way connected through symmetric low-depth local quantum channels.

The rest of this paper is organized as follows. In Sec.~\ref{Sec: def}, we propose formal definitions of SW-SSB and establish a few basic universal properties. In particular, we prove that SW-SSB, as in Def. \ref{def:swssb}, is robust under low-depth local channels, and thus can be used as a universal characterization of mixed-state quantum phases. In Sec.~\ref{sec:thermal} we discuss SW-SSB for thermal states in the canonical ensemble at nonzero temperatures. We then explore a number of examples of SW-SSB transitions in decohered/noisy many-body states in Secs.~\ref{sec:Ising} and \ref{sec:rotor}. In Sec.~\ref{sec:recover} we discuss the recoverability of the decohered Ising model. We show that as long as two states belong to the same phase (both with or without the SW-SSB order), they are two-way connected through log-depth symmetric local quantum channels. We end with a summary and outlook in Sec. \ref{Sec: summary}. Some peripheral details are presented in the Appendices.

\section{Definition and Properties}
\label{Sec: def}
In this section, we give two inequivalent definitions of SW-SSB order parameters: fidelity correlator and R\'enyi-2 correlator of local operators carrying strong symmetry charge. Then we discuss some characteristic properties of SW-SSB mixed-state. {We will see that, while the R\'enyi-2 correlator is in general simpler to calculate, the fidelity correlator offers a more fundamental characterization of the universal properties of SW-SSB.}

\subsection{Fidelity correlator and stability theorem}
\label{Sec:fidelity}

Recall that the fidelity\footnote{In literature, the fidelity is often defined as the square of Eq.~\eqref{eq:Fdef}. For our purpose it is more natural to adopt Eq.~\eqref{eq:Fdef}.} between two states $\rho$ and $\sigma$ is defined as
\begin{equation}
\label{eq:Fdef}
    F(\rho, \sigma)=\Tr \sqrt{\sqrt{\rho}\sigma\sqrt{\rho}}.
\end{equation}
When both $\rho$ and $\sigma$ are pure states, $F(\rho,\sigma)$ becomes the modulus of their overlap. Fidelity is a commonly used measure of distance between mixed states. Motivated by fidelity, we define the \emph{fidelity} average of an operator $O$ in a mixed state $\rho$ as follows:
\begin{equation}
    \braket{O}_F = F(\rho, O\rho O^\dag).
\end{equation}
Again, if $\rho=\ketbra{\psi}{\psi}$, then $\braket{O}_F=\left|\braket{\psi|O|\psi}\right|$, the modulus of the expectation value. But for a generic mixed state, the fidelity average is very different from the usual expectation value defined as $\braket{O}=\Tr[\rho O]$. If $O$ is unitary, then the above expression is the actual fidelity between two different density matrices $\rho$ and $\sigma:=O\rho O^\dag$. {If $O$ is charged under a strong symmetry of $\rho$, then $\langle O\rangle_F=0$.} {\color{black}Notice that
\begin{eqnarray}
    \langle O\rangle_F&=&\Tr\sqrt{\sqrt{\rho}O\rho O^{\dagger}\sqrt{\rho}}=\Vert \sqrt{\rho}O\sqrt{\rho}\Vert_1 \nn
    &>&|\Tr\sqrt{\rho}O\sqrt{\rho}|=|\langle O\rangle|,
\end{eqnarray}
which implies that if $\langle O\rangle\neq0$, then we must have $\langle O\rangle_F>0$.}

For a local operator $O(x)$, define the \emph{fidelity correlator} as the fidelity average of $O(x)O^\dag(y)$, or more explicitly:
\begin{equation}
F_O(x,y)=\braket{O(x)O^\dag(y)}_F.
\label{fidelity correlator}
\end{equation}

\begin{definition}\label{def:swssb}
A mixed state $\rho$ with a strong symmetry $G$ has \textbf{strong-to-weak SSB} (SW-SSB) if the fidelity correlator is long-range ordered:
\begin{equation}
\label{eq:sqrtR2}
\lim_{|x-y| \to \infty} F_O(x,y) = c > 0,
\end{equation}
for some charged local operator $O(x)$, and at the same time there is no long-range order in the usual sense, i.e. for any charged local operator $O'$,
 $\langle O'(x)O'^\dag(y) \rangle_\rho \rightarrow0$ as $|x-y|\to\infty$.  
\end{definition}
{Hereinafter, we will denote $\lim_{|x-y| \to \infty} F_O(x, y) = c > 0$ by $F_O(x,y) \sim O(1)$, and similarly to other correlators.}

The simplest case of a charged local operator $O(x)$ is when it transforms under the symmetry group $G$ as $g O(x) g^{-1} = \lambda_g O(x)$, for some $\lambda_g \in \U$. If $O_a(x)$ is part of a higher-dimensional unitary representation of $G$, such as the spin operators $S^{(\alpha)}$ of SO(3), then we can replace $O(x) O^\dagger(y)$ by the quadratic symmetry-invariant operator $\sum_a O_a(x) O^\dagger_a(y)$.

\rm We now discuss the simplest example in an Ising spin system. The prototypical state with SW-SSB is $\rho_0\propto \mathbbm{1}+X$ with a strong $\Z_2$ symmetry generated by $X\equiv\prod_iX_i$. $\rho_0$ is analogous to the GHZ cat state $\ket{00\cdots} + \ket{11 \cdots}$ of the usual $\Z_2$ SSB. In particular, if we write $\rho_0$ in the $Z$-basis, it is a convex sum of cat states:
\begin{align}
    \rho_0\propto \sum\limits_{s}(|s\rangle+X|s\rangle)(\langle s|+\langle s|X),
\end{align}
where $s \in \{0,1\}^N$ labels the bit string of spins in the $Z$-basis with $N$ being the total number of spins. It is easy to see that the fidelity correlator $F_{Z}(x,y)$ for the order parameter $Z(x)$ is exactly equal to 1, independent of $x$ and $y$.

More generally, let us assume $\rho$ is a mixture of orthogonal pure states: $\rho=\sum_a p_a\ketbra{\psi_a}{\psi_a}$, such that $\braket{\psi_a|O(x)O^\dag(y)|\psi_b}\propto\delta_{ab}$.  Then we can write the fidelity correlator as 
\begin{equation}
    {\sum_a p_a \left|\braket{\psi_a |O(x)O^\dag(y)|\psi_a}\right|   }\equiv\overline{\left|\braket{O(x)O^\dag(y)}\right|}.
\end{equation}
This is nothing but the Edward-Anderson correlator used for spin glass order if $\rho$ is an ensemble of SSB states. Recall that in a disordered ensemble, the Edward-Anderson order parameter measures exactly the spontaneous breaking of an exact symmetry down to an average symmetry. {From this expression we can also see that SW-SSB is only applicable to mixed states since the fidelity correlator of pure states is just the (absolute value of) linear correlator, which we assume is not long-range ordered.} We summarize the key patterns of different mixed-state SSB in Table~\ref{tab:SSBpatterns}.

\begin{table}
\renewcommand\arraystretch{1.4}
\centering
\begin{tabular}{|c|c|c|}
\hline
Symmetry & $F_O(x,y)$ & $\Tr(\rho O(x)O^\dag(y))$ \\
\hline
Unbroken & 0 & 0\\
\hline
SW-SSB  & $O(1)$ & 0 \\
\hline
Completely Broken & $O(1)$ & $O(1)$\\
\hline
\end{tabular}
\caption{The behavior of SSB diagnostics in a mixed state with a strong symmetry. The operator $O$ is chosen as a generic local operator transformed in a nontrivial representation of the strong symmetry. 0 denotes exponential decay with respect to the distance $|x-y|$.}
\label{tab:SSBpatterns}
\end{table}

We now study the robustness of SW-SSB as we deform our states using low-depth quantum channels. We shall start by defining such channels:

{
\definition A quantum channel $\mathcal{E}$ is termed a symmetric low-depth channel if it can be purified to a symmetric local unitary operator $U$ acting on $\mathcal{H} \otimes \mathcal{H}^a$, where $\mathcal{H}$ and $\mathcal{H}^a$ are the physical and ancilla Hilbert spaces, respectively. The channel operation $\mathcal{E}(\rho)$ is given by $\Tr_a[U^\dagger (\rho \otimes | 0 \rangle \langle 0 | )U]$, where $| 0 \rangle$ represents a product state in $\mathcal{H}^a$. Specifically:
\begin{enumerate}[1.]
\item $\mathcal{H}^a = \otimes_i \mathcal{H}_i^a$ is a tensor product of local Hilbert spaces on each site of $\mathcal{H}$. $\dim(\mathcal{H}_i^a)$ is finite for each site.
\item The unitary circuit $U$ has low depth $D$. In particular, the channel is called finite-depth if $D\sim O(1)$, and logarithmic depth if  $D\sim {\rm PolyLog}(L)$, where $L$ represents the linear size of the system. 
\item The channel is \textit{strongly symmetric} under the symmetry $g$ if each gate of the purified unitary $U$ commutes with $\tilde{g}\equiv g\otimes I^a$, where $I^a$ is the identity operator on the ancilla space.
\end{enumerate} 
}

The following \textbf{stability theorems} establishes SW-SSB as a universal property of mixed-state quantum phases:

\begin{theorem}\label{thm:SW-SSB_stability}
If a mixed state $\rho$ has SW-SSB (Eq.~\eqref{eq:sqrtR2})  and $\E_{\mathrm{SFD}}$ is a strongly symmetric finite-depth (SFD) local quantum channel, then $\E_{\mathrm{SFD}}[\rho]$ will also have SW-SSB. 
\end{theorem}

\begin{proof}
\rm Recall Uhlmann's theorem \cite{Uhlmann} that fidelity can be defined in terms of purification as
\begin{equation}
F(\rho,\sigma)=\max_{|\psi_\rho\rangle,|\phi_\sigma\rangle}|\langle\psi_{\rho}|\phi_{\sigma}\rangle|,
\label{Uhlmann}
\end{equation}
where $|\psi_{\rho}\rangle, |\phi_{\sigma}\rangle$ run over all possible purifications of $\rho,\sigma$ respectively, and the maximum is taken with respect to all possible purifications. The form of $\sigma=O\rho O^{\dagger}$ ($O\equiv O(x)O^{\dagger}(y)$, $O(x)$ and $O(y)$ are charged operators) means that the fidelity can be rewritten as
\begin{equation}
F(\rho,\sigma)=\max_{|\psi_\rho\rangle,|\phi_\rho\rangle}|\langle\psi_{\rho}|O|\phi_{\rho}\rangle|.
\label{Uhlmann1}
\end{equation}
Now we suppose $F(\rho,\sigma)\sim O(1)$ for the mixed state $\rho$, and consider an SFD channel $\E$, which can be purified to a symmetric finite-depth local unitary operator $U$ acting on the physical system tensor a trivial product state $|0\rangle\langle0|_a$ in an ancilla Hilbert space $\mathcal{H}^a$. Gathering everything together, we have
\begin{align}
O(1)&\sim|\langle\psi_\rho|O|\phi_\rho\rangle|=|(\langle\psi_{\rho}|\otimes \langle 0|_a)O(|\phi_{\rho}\rangle\otimes|0\rangle_a)| \nn
&=|(\langle\psi_{\rho}|\otimes \langle 0|_a)U^{\dagger}UOU^{\dagger}U(|\phi_{\rho}\rangle\otimes|0\rangle_a)| \nn
&=|\langle\psi_{\E[\rho]}|O'|\phi_{\E[\rho]}\rangle|,
\label{Uhlmann2}
\end{align}
where $O'=UOU^{\dagger}=O'(x)O'^{\dagger}(y)$, and by the SFD nature of $U$, $O'(x), O'(y)$ are local and carry the same symmetry charge as $O(x),O(y)$. But $O'$ will now act in the ancilla space as well. Decompose $ O'=\sum_i O_i\otimes O_i^a$ where $O_i$ acts only in the physical Hilbert space and $O_i^a$ acts only in the ancilla Hilbert space $\mathcal{H}^a$. Crucially, the SFD nature of $U$ means there are only finitely many terms in the decomposition. So there must be at least one $i$ such that
\begin{align}
\label{eq:fidproof}
O(1)&\sim|\langle\psi_{\E[\rho]}|O_i\otimes O_i^a|\phi_{\E[\rho]}\rangle| \nn
    &\leq \Vert O_i^a\Vert\cdot |\langle\psi_{\E[\rho]}|O_i\otimes U_a|\phi_{\E[\rho]}\rangle| \nn
    &\leq \Vert O_i^a\Vert\cdot \max_{\rm purification}|\langle\psi'_{\E[\rho]}|O_i|\phi'_{\E[\rho]}\rangle| \nn
    &=\Vert O_i^a\Vert\cdot F\left(\E[\rho],
    O_i\E[\rho]O^{\dagger}_i\right),
\end{align}
where $U_a$ is some unitary acting only in the ancilla Hilbert space $\mathcal{H}^a$, and $\Vert O_i^a\Vert$ is the operator norm of $O_i^a$, which is crucially finite. The third line comes from the fact that $U_a|\phi_{\E[\rho]}\rangle$ is just another purification of $\E[\rho]$. So we conclude that the fidelity correlator for $O_i$ on the state $\E[\rho]$ is $O(1)$. By the strong symmetry condition, $O_i$ carries the same charge as $O$. So $\E[\rho]$ also has SW-SSB.
\end{proof}

Theorem \ref{thm:SW-SSB_stability} can be extended to more general low-depth channels:

\begin{theorem} \label{thm:SW-SSB_stability2}
If a mixed state $\rho$ has SW-SSB (Eq.~\eqref{eq:sqrtR2}) with $F_O(x,y)\sim O(1)$ for some charged operators $O(x)$, $O(y)$, and $\E_{\mathrm{SFD}}$ is a strongly symmetric depth-$D$ local quantum channel, then $\E_{\mathrm{SFD}}[\rho]$ will have $F_{\tilde{O}}(x,y)\sim \exp (-{\rm Poly}(D))$ for some charged local operators $\tilde{O}(x), \tilde{O}(y)$. In particular, for $D\sim {\rm PolyLog}(L)$, $F_{\tilde{O}}$ decays sub-exponentially with system size $L$.
\end{theorem}
\begin{proof}
\rm We follow the same steps as the proof of Theorem~\ref{thm:SW-SSB_stability}  till Eq.~\eqref{Uhlmann2}. Now the evolved operator $O'\equiv U OU^{\dagger}$ acts nontrivially on two regions $X, Y$ (around the points $x,y$) with size $|X|,|Y|\sim D^d$, where $d$ is the spatial dimension. We can now decompose the operator $O'$ as
\begin{equation}
    O' = \sum_{i} c_{ik} O_i\otimes O_k^a,
\end{equation}
where $\{O_i=O_i(x)O^{\dagger}_i(y)\}$ and $\{O^a_i=O^a_i(x)(O^a_i)^{\dagger}(y)\}$ are orthogonal families with respect to the Hilbert-Schmidt inner product. We can further choose $\{O^a_k\}$ to be unitary, and $\{O_i\}$ to have the normalization $\tr_{\mathcal{H}} (O_i^{\dagger}O_j) =\delta_{ij}\tr_{\mathcal{H}} (O^{\dagger}O)$. Recall that $O'=U O U^{\dagger}$ has the same Hilbert-Schmidt norm as $O\otimes I^a$, so the above normalization requires that $\sum_{ik}|c_{ik}|^2=1$. We can now proceed with a refined version of Eq.~\eqref{eq:fidproof}:
\begin{eqnarray}
    O(1)&\sim& |\sum_{ik}c_{ik}\langle \psi_{\mathcal{E}[\rho]}|O_i\otimes O_k^a|\phi_{\mathcal{E}[\rho]}\rangle| \nn 
    &\leq& \sum_{ik} \Vert c_{ik} O_k^a \Vert \cdot F\left(\E[\rho],
    O_i\E[\rho]O^{\dagger}_i\right), \nn
    &\leq& \left(\sum_{ik} |c_{ik}| \right) \max_i F\left(\E[\rho],
    O_i\E[\rho]O^{\dagger}_i\right).
\end{eqnarray}

Finally, since $\sum_{ij}|c_{ij}|^2=1$, the expression $\sum_{ij}|c_{ij}|$ is upper bounded by the square root of the number of terms in the summation, which is $\sim \exp({\rm Poly}(|X|,|Y|))\sim \exp ({\rm Poly}(D))$. We therefore conclude that for some operator $O_i\equiv \tilde{O}$ acting on the system Hilbert space $\mathcal{H}$, we have
\begin{equation}
    F\left(\E[\rho],
    \tilde{O}\E[\rho]\tilde{O}^{\dagger}\right)\geq \exp(-{\rm Poly}(D)).
\end{equation}
\end{proof}

{ We note that the stability theorems are only concerned with the fidelity correlator, and are not sensitive to the ordinary linear correlator. In fact, the linear correlator in ordinary SSB need not be stable against symmetric finite-depth channels. As a simple example, consider a one-dimensional GHZ state $|\uparrow\uparrow...\uparrow\rangle+|\downarrow\downarrow...\downarrow\rangle$, and perform strong measurements for all the $X_i$ operators. The resulting state is $\one+X$: the strong-to-trivial SSB (measured by ordinary correlator) is eliminated by the strong measurements, but the strong-to-weak SSB survives.
}

The stability theorem implies that a mixed state $\rho$ with SW-SSB cannot be brought to a symmetric pure product state $|00\cdots\rangle$, using a symmetric low-depth channel $\mathcal{E}$, since the latter has $F_O(x,y)=0$ for any charged operator $O$. We can make the statement slightly stronger: the state $\rho$ is not only non-trivializable but also \textit{non-invertible}. First let us define the notion of invertible states:

\definition   A mixed state on the physical Hilbert space $\rho\in\HH$ is \emph{symmetrically invertible} if there exists an ancillary mixed states $\tilde{\rho} \in \tilde{\HH}$ (with the symmetry element $g$ acting as $U_g\otimes \tilde{U}_g$), and a pair of symmetric low-depth local channels $\E_T, \E_P : Q(\HH \otimes \tilde{\HH}) \to Q(\HH \otimes \tilde{\HH})$ such that
    \begin{equation}
    \label{eq:invertible}
        \rho \otimes \tilde{\rho} \xrightarrow{\E_T} |00\cdots\rangle\langle00\cdots| \xrightarrow{\E_P} \rho \otimes \tilde{\rho},
    \end{equation}
    where $|00\cdots\rangle \in \HH \otimes \tilde{\HH}$ is a symmetric pure product state. 

\rm

The above definition naturally generalizes the important notion of invertibility in pure states. It has been shown that the notion of invertibility is also important for discussing topological phases in mixed states, in the sense that (1) symmetry-protected (SPT) topological phases in mixed states are well-defined only for (bulk) invertible states \cite{Ma:2023rji}; and (2) systems with anomalous symmetry (such as those on the boundaries of nontrivial SPT states) cannot be symmetrically invertible \cite{lessa2024mixedstate}.

The stability theorems immediately imply that density matrices with SW-SSB are not symmetrically invertible:

\begin{corollary}
A density matrix $\rho\in\mathcal{H}$ with SW-SSB is not symmetrically invertible.
\end{corollary}
\begin{proof}
Suppose the (partial) contrary: $\exists\Tilde{\rho}$ and a strongly symmetric low-depth quantum channel $\E$, such that 
\begin{align}
\E[\rho\otimes\Tilde{\rho}]=|00\cdots\rangle\langle00\cdots|.
\end{align}
Since the pure product state has $F_O(x,y)=0$ for any local charged operator $O$, the stability theorem requires $\rho\otimes \tilde{\rho}$ to also have $F_O(x,y)=0$ for $|x-y|$ much larger than the channel depth $D$ (or exponentially small in $|x-y|$ if the gates in the channel have exponential tails). In particular, for any local charged operator $O_s\equiv O_s(x)O_s^{\dagger}(y)$ that only acts on the original system $\HH$,
\begin{equation}
    F(\rho,O_s\rho O_s^{\dagger})=F(\rho\otimes\tilde{\rho},O_s\rho\otimes\tilde{\rho} O_s^{\dagger})\to 0,
\end{equation}
which means the original state $\rho$ cannot have SW-SSB.
 
\end{proof}

We now comment on some alternative definitions of SW-SSB. We note that physically the fidelity defined in Eq. \eqref{eq:sqrtR2} measures how distinguishable $\rho$ and $\sigma$ are, where $\sigma\equiv O(x)O^{\dagger}(y)\rho O(y)O^{\dagger}(x)$.
In fact, many other information-theoretic quantities measure the similarity of two mixed states. For example, consider the quantum relative entropy that is defined as
\begin{align}
S(\rho\Vert\sigma)=\Tr(\rho\log\rho-\rho\log\sigma).
\end{align}
The trace distance is defined as
\begin{align}
D(\rho,\sigma)=\frac{1}{2}\Vert\rho-\sigma\Vert_1.
\end{align}
Analogous to Eq. \eqref{eq:sqrtR2}, one can define SW-SSB based on these measures: a state $\rho$ has SW-SSB when the similarity between $\rho$ and $\sigma$ is non-vanishing and does not depend on $|x-y|$ at large distances. For example, SW-SSB in the sense of quantum relative entropy means $S(\rho\Vert\sigma)$ is upper bounded as $|x-y|\rightarrow \infty$.

All these distinguishability measures (fidelity, quantum relative entropy, and trace distance among others) share several properties that make them proper notions of distance between quantum states~\cite{nielsen00}. Most importantly for our purpose, they satisfy the property known as ``data processing inequality". It says that the distance between two states is monotonic under quantum channels, such that the two states become less distinguishable. Specifically, for two states $\rho$ and $\sigma$ and a quantum channel $\mathcal{E}$, we have 
\begin{equation}
    F(\rho, \sigma)\leq F\big(\mathcal{E}[\rho], \mathcal{E}[\sigma]\big).
\end{equation}
Similar inequalities hold for quantum relative entropy and trace distance. This property directly gives a stability theorem when the order parameter operator $O$ commutes with the quantum channel.

A natural question arises regarding the relationship between SW-SSBs defined in terms of the distinguishability measures stated above. Based on the Fuchs-van de Graaf inequality~\cite{nielsen00}:
\begin{align}
1-F(\rho,\sigma)\leq D(\rho,\sigma)\leq\sqrt{1-F^2(\rho,\sigma)},
\label{Fuchs-van de Graaf}
\end{align}
the fidelity and trace distance definitions of SW-SSB are entirely equivalent. 

Next, we show that  if a state $\rho$ has SW-SSB defined in terms of quantum relative entropy, it must also have SW-SSB defined by fidelity. To this end and for later use, we introduce the sandwiched R\'enyi relative entropy, defined as:
\begin{equation}
    \Tilde{D}_\alpha(\rho\Vert \sigma) = \frac{1}{\alpha-1}\log_2 \Tr [(\sigma^\frac{1-\alpha}{2\alpha} \rho \sigma^\frac{1-\alpha}{2\alpha})^\alpha].
    \label{eq:sandwiched}
\end{equation}
This assertion that SW-SSB defined by quantum relative entropy implies SW-SSB defined by fidelity relies on three properties of $\Tilde{D}_\alpha$~\cite{Muller-Lennert:2013liu}: 
\begin{enumerate}
    \item  $\Tilde{D}_\alpha(\rho \Vert \sigma)$ converges to the fidelity correlator in Eq. \eqref{eq:sqrtR2} for $\alpha\to \frac{1}{2}$: $\Tilde{D}_\frac{1}{2}(\rho\Vert\sigma) = -2\log_2 F(\rho,\sigma)$.
    \item $\Tilde{D}_\alpha(\rho \Vert \sigma)$ converges to the quantum relative entropy $S(\rho \Vert \sigma)$ for $\alpha\to 1$;
    \item For $0<\alpha<\beta<1$, $\Tilde{D}_\alpha(\rho\Vert \sigma) \leq \Tilde{D}_\beta(\rho\Vert \sigma)$
\end{enumerate}
As a result, we have $F(\rho,\sigma)\geq e^{-\frac{\ln 2}{2}S(\rho\Vert\sigma)}$. A SW-SSB defined in terms of quantum relative entropy thus implies that the fidelity correlator has an $O(1)$ expectation at large $|x-y|$. In fact, as a good distinguishability measure (with data-processing inequality), the sandwiched R\'enyi relative entropy can also be used to define SW-SSB.

We conjecture that all the above definitions are equivalent and will provide supporting evidence in Sec. \ref{sec:example}.

\subsection{R\'enyi-2 correlator}
The similarity measures of two mixed states are not limited to the above information-theoretic quantities. In this subsection, we introduce an alternative definition of SW-SSB that is not equivalent to the definitions in Sec. \ref{Sec:fidelity}.

A useful way to study the mixed quantum state is utilizing the Choi–Jamiołkowski isomorphism. It maps a density matrix $\rho$ to a pure state $|\rho\rrangle$ in the doubled Hilbert space. More explicitly, the Choi–Jamiołkowski isomorphism of a density matrix $\rho=\sum\lambda_j|\psi_j\rangle\langle\psi_j|$ is the following Choi state in the doubled Hilbert space $\mathcal{H}_d=\mathcal{H}_u\otimes\mathcal{H}_l$,
\begin{align}
|\rho\rrangle=\frac{1}{\sqrt{\mathrm{dim}(\rho)}}\sum\limits_jp_j|\psi_j\rangle\otimes|\psi_j^*\rangle,
\label{Choi state}
\end{align}
where both upper Hilbert space $\mathcal{H}_u$ and lower Hilbert space $\mathcal{H}_l$ are copies of the physical Hilbert space $\mathcal{H}$. The operator $\sigma\equiv O(x)O^{\dagger}(y)\rho O(y)O^{\dagger}(x)$ is mapped to the following state in the doubled Hilbert space:
\begin{align}
|\sigma\rrangle=O_u(x)O_u^\dag(y)O_l(y)O_l^\dag(x)|\rho\rrangle,
\end{align}
where $O_u$ and $O_l$ are charged operators defined in the upper and lower Hilbert spaces, respectively. The strong symmetry $G$ is mapped to a double symmetry $G_u\times G_l$, where $O_u/O_l$ only carries charge of $G_u/G_l$.

The natural similarity measure of $\rho$ and $\sigma$ in the doubled Hilbert space $\mathcal{H}_d$ is simply the overlap between $|\rho\rrangle$ and $|\sigma\rrangle$ (with a normalization), namely
\begin{align}
\frac{\llangle\rho|\sigma\rrangle}{\llangle\rho|\rho\rrangle}=\frac{\llangle\rho|O_u(x)O_u^\dag(y)O_l(y)O_l^\dag(x)|\rho\rrangle}{\llangle\rho|\rho\rrangle}.
\label{double correlator}
\end{align}
This correlator precisely measures the SSB in doubled Hilbert space that breaks the double symmetry $G_u\times G_l$ to its diagonal subgroup symmetry~\cite{Lee_2023}.

Mapping the correlator \eqref{double correlator} back to the physical Hilbert space, we obtain the following R\'enyi-2 correlation function of the charged operator $O(x)$, namely
\begin{equation}
R^{(2)}(x,y):=\frac{{\Tr}\left(O(x)O^{\dagger}(y)\rho O(y)O^{\dagger}(x)\rho \right)}{{\Tr}\rho^2}.
\end{equation}
In particular, the doubled symmetry $G_u\times G_l$ and its diagonal subgroup symmetry in the doubled Hilbert space are mapped to the strong and weak symmetry in the physical Hilbert space, respectively. 

Notice that for the kind of ``spin glass" ensemble discussed in Sec. \ref{Sec:fidelity}, the R\'enyi-2 correlator becomes
\begin{equation}
    R^{(2)}(x,y)\approx \frac{\sum_a p_a^2 \braket{\psi_a|Z_xZ_y|\psi_a}^2}{\sum_a p_a^2}.
\end{equation}

Gathering everything together, we formulate an alternative definition of SW-SSB:

\begin{definition}\label{def:renyi_ssb}
A mixed state $\rho$ with strong symmetry $G$ has \textbf{strong-to-weak SSB in the sense of R\'enyi-2 correlator} if the R\'enyi-$2$ correlator of some charged local operator $O(x)$ is non-vanishing:
\begin{equation}
\label{eq:R2}
\lim\limits_{|x-y|\to\infty}\frac{{\Tr}\left(O(x)O^{\dagger}(y)\rho O(y)O^{\dagger}(x)\rho \right)}{{\Tr}\rho^2}\sim O(1),
\end{equation}
while the conventional correlation function shows no long-range order:
\begin{align}
\lim\limits_{|x-y|\rightarrow\infty}{\Tr}\left(\rho O(x)O^\dag(y)\right)=0.
\end{align}
\end{definition}
For the simplest example $\rho_0\propto\mathbbm{1}+X$ with a strong $\Z_2$ symmetry, it is easy to see that the R\'enyi-2 correlator of the order parameter $Z$ is exactly 1, independent of the distance between $x$ and $y$. 

We now demonstrate that two definitions of SW-SSB through fidelity correlator \eqref{eq:sqrtR2} and R\'enyi-2 correlator \eqref{eq:R2} are \emph{inequivalent} through a simple example. We consider the following state in an Ising spin chain
\begin{equation}
    \rho=\frac{1}{2}|\!++\cdots\rangle\langle ++\cdots|+\frac{1}{2^{L+1}}(\one+X),
\end{equation}
which is a convex sum of an ideal SW-SSB state $\one+X$ and a symmetric pure product state $|++\cdots\rangle$. For the order parameter $Z$, we have
\begin{align}
    \sigma \equiv & Z(x)Z(y)\rho Z(x)Z(y) \nonumber \\
    = & \frac{1}{2}Z(x)Z(y)|\!++\cdots\rangle\langle ++\cdots|Z(x)Z(y) + \nonumber \\
      &+\frac{1}{2^{L+1}}(\one+X).
\end{align}

It is then straightforward to check that $F_O(x,y)\to O(1)$ while $R^{(2)}(x,y)\to 0$, namely the state $\rho$ has SW-SSB in the fidelity sense, but not in the R\'enyi-$2$ sense. { The opposite situation, in which $F_O(x, y) \to 0$ but $R^{(2)}(x, y) \to O(1) > 0$, can also happen, as shown in Appendix~\ref{app:counterexample}.} Another example illustrating the inequivalence between the fidelity correlator and the R\'enyi-2 correlator will be provided in Section \ref{sec:example}, where these two correlators are mapped to observables in distinct statistical models.

The major weakness of the R\'enyi-2 correlator definition, compared to the one based on fidelity, is the lack of a stability theorem (such as Thm.~\ref{thm:SW-SSB_stability} and \ref{thm:SW-SSB_stability2}). In fact, we will discuss an explicit example in Sec.~\ref{sec:recover} where the broken symmetry in the R\'enyi-2 sense is restored using a low-depth symmetric channel.  

Although a stability theorem does not exist for R\'enyi-2 SW-SSB, we can prove that the SW-SSB defined through the R\'enyi-2 correlator Eq.~\eqref{eq:R2} 
shares the same property of \textit{non-invertibility} with the SW-SSB defined through the fidelity Eq.~\eqref{eq:sqrtR2}.

\begin{theorem}
\label{thm:noninvertible}
A mixed state $\rho$ with SW-SSB in the R\'enyi-2 correlator sense Eq.~\eqref{eq:R2} is not symmetrically invertible.
\end{theorem}
\begin{proof}
Suppose the (partial) contrary: $\exists\Tilde{\rho}\in\widetilde{\mathcal{H}}$ and a strongly symmetric local quantum channel $\E$, such that
\begin{align}
\E[\rho\otimes\Tilde{\rho}]=|0\rangle\langle0|\in\mathcal{H}\otimes\widetilde{\mathcal{H}}.
\end{align}
We purify the channel $\E$ with an ancilla Hilbert space $\mathcal{H}^a$ to a low-depth unitary $U$, namely
\begin{align}
U\left(\rho\otimes\Tilde{\rho}\otimes|0\rangle\langle0|_a\right)U^\dag=|0\rangle\langle0|\otimes\rho_a,
\end{align}
where $\rho_a$ is a density matrix in $\mathcal{H}^a$. Then the R\'enyi-2 correlator in the state $\E(\rho\otimes \tilde{\rho})$ is given by
\begin{align}
\frac{\tr\left[O_x'O_y'^\dag(|0\rangle\langle0|\otimes\rho_a)O_y'O_x'^\dag(|0\rangle\langle0|\otimes\rho_a)\right]}{\tr(\rho_a^2)},
\label{eq:invR2}
\end{align}
where $O_{x}'=UO(x)U^\dag$ and $O_{y}'=UO(y)U^\dag$. Recall that a unitary does not change the purity of a density matrix. Now because $\E$ is strongly symmetric (and so is the purification $U$), $O'$ carries the same charge of strong symmetry as $O$. Expand $O_x'$ as the following form
\begin{align}
O_x'=\sum\limits_io_i(x)\otimes o_i^a(x),
\end{align}
where $o_i^a(x)$ acts only on the ancilla $\mathcal{H}^a$ and $o_i(x)$ acts on $\mathcal{H}\otimes\widetilde{\mathcal{H}}$, and the charge of strong symmetry is carried by $o_i$ only. Then the numerator of Eq. \eqref{eq:invR2} is a sum of terms involving the factor $\langle00\cdots|o_i^\dag(x)o_j(y)|00\cdots\rangle\sim e^{-|x-y|/\xi}$ (or strictly zero if the gates do not have exponential tails). Therefore, we conclude that the R\'enyi-$2$ correlator has to be exponentially small for invertible states. In particular, the original $\rho$ cannot have SW-SSB in the R\'enyi-$2$ sense.
\end{proof}

In particular, the above theorem implies that under a symmetric shallow channel, even though the R\'enyi-$2$ long-range order may be destroyed, the state cannot be brought to a completely trivial state (a symmetric pure product state) -- namely some kind of nontrivial order (for examples SW-SSB in the fidelity sense) should persist. Again this will be demonstrated by the explicit example in Sec.~\ref{sec:example} and \ref{sec:recover}. 

To summarize, the overall lessons are
\begin{enumerate}[1.]
    \item The R\'enyi-$2$ correlator, as a characterization of SW-SSB, is not as universal as the fidelity correlator. In particular, the Stability Theorems~\ref{thm:SW-SSB_stability},\ref{thm:SW-SSB_stability2} does not hold for the R\'enyi-2 correlator, as will be exemplified in Sec. \ref{sec:Ising};
    \item Nevertheless, having a non-vanishing R\'enyi-$2$ correlator still indicates the nontriviality of a state, due to the Non-invertibility Theorem~\ref{thm:noninvertible}. The R\'enyi-$2$ correlator remains valuable for this reason, especially given that the R\'enyi-$2$ correlator is in general much simpler to calculate than the fidelity correlator.
    {
    \item \color{black}Experimentally measuring the fidelity correlator is extremely hard in the sense that it generally calls for the full quantum state tomography (QST) of the density matrix, whose complexity is exponential with respect to the system size. Therefore, the other significance of introducing the R\'enyi-2 correlator is the experimental accessibility: the R\'enyi-2 correlator can be experimentally measured without performing QST. Recently, the proposed \textit{classical shadow tomography} \cite{Huang_2020, Elben_2022, Hu_2023} significantly reduces the experimental complexity of measuring nonlinear quantum information-theoretic quantities, especially in open quantum systems. In particular, as indicated in Eq. \eqref{eq:R2}, both the numerator and denominator of the R\'enyi-2 correlator are polynomials of the density matrix $\rho$, which emphasizes that both can be efficiently measured by classical shadow tomography.
    }
\end{enumerate}

\subsection{Local indistinguishability and detectability}
Recall that pure-state SSB can also be characterized through degenerate states that are locally indistinguishable. For concreteness, consider the following two orthogonal $\Z_2$ SSB states, with different symmetry charges:
\begin{align}
|\psi_\pm\rangle=\frac{1}{\sqrt{2}}\left(|\!\uparrow\cdots\uparrow\rangle\pm|\!\downarrow\cdots\downarrow\rangle\right).
\end{align}
For any $k$-point ($k$ finite) observable in a simply connected region
\begin{align}
\label{eq:kpt}
M_k\equiv M(x_1)M(x_2)\cdots M(x_k),
\end{align}
the two states give identical expectation values:
\begin{align}
\langle\psi_-|M_k|\psi_-\rangle\simeq\langle\psi_+|M_k|\psi_+\rangle.
\end{align}
In other words, $|\psi_{\pm}\rangle$ are \textit{locally indistinguishable}. Furthermore, if we make a superposition, for example $|\psi\rangle=\frac{1}{\sqrt{2}}(|\psi_+\rangle+|\psi_-\rangle)=|\!\!\uparrow\uparrow\cdots\rangle$, then $|\psi\rangle$ explicitly breaks the symmetry, in a way that is locally detectable: $\langle\psi|Z(x)|\psi\rangle=1$.

Below we demonstrate that a similar notion of local indistinguishability exists for mixed states with SW-SSB. 

For a density matrix $\rho$ with SW-SSB, we construct a new state 
\begin{align}
\label{eq:indisrho'}
\tilde\rho\equiv\frac{1}{L^d}\sum\limits_xO(x)\rho O^\dag(x),
\end{align}
here we suppose $O(x)$ is a unitary charged operator to ensure that $\tilde\rho$ is a legitimate density matrix, for simplicity.
The presence of strong symmetry implies that $\rho$ and $\tilde\rho$ are orthogonal in the fidelity sense: $F(\rho,\tilde\rho)=0$ because they carry different symmetry charges. 


For any $k$-point observable (Eq.~\eqref{eq:kpt}), we have the following simple relations:
\begin{eqnarray}
{\Tr}[M_k(\tilde{\rho}-\rho)]&=&\frac{1}{L^d}{\Tr}\left[M_k\sum_x\left(O(x)\rho O^{\dagger}(x)-\rho \right) \right] \nn
&=&\frac{1}{L^d}\sum_x{\Tr}\left( [M_k,O(x)]\rho O^{\dagger}(x)\right) \nn
&=&O(k/L^d),
\label{eq:localindistin}
\end{eqnarray}
{
\color{black}
and 
\begin{widetext}
\begin{align}
&F\left(\rho, M_k \rho M_k^\dag\right)-F\left(\tilde\rho, M_k \tilde\rho M_k^\dag\right) \leq F\left(\rho, M_k \rho M_k^\dag\right)-\frac{1}{L^d}\sum_xF\left(O_x \rho O_x^\dag, M_k O_x \rho O_x^\dag M_k^\dag\right)\nonumber\\
=& F\left(\rho, M_k \rho M_k^\dag\right)-\frac{1}{L^d}\sum_xF\left(O_x \rho O_x^\dag, \left([M_k,O_x]+O_xM_k\right)\rho\left(M_k^\dag O_x^\dag + \left[O_x^\dag, M_k^\dag\right]\right)\right)\nonumber\\
=& \frac{1}{L^d}\sum_{x\in k}F\left(\rho, M_k \rho M_k^\dag\right)-F\left(O_x \rho O_x^\dag, \left([M_k,O_x]+O_xM_k\right)\rho\left(M_k^\dag O_x^\dag + \left[O_x^\dag, M_k^\dag\right]\right)\right)\sim O\left(\frac{k}{L^d}\right),
\end{align}
\end{widetext}
where $x\in k$ means that the site $x$ locates inside the support of $M_k$. The inequalities above came from the joint concavity of fidelity: 
\begin{equation}
F(\sum_ip_i\rho_i,\sum_ip_i\sigma_i)\geq \sum_ip_iF(\rho_i,\sigma_i).
\label{Eq: joint concavity}
\end{equation}
}
We conclude that any finite-point observable cannot distinguish $\tilde{\rho}$ from $\rho$. To distinguish the two states, the observable $M$ needs to act nontrivially on the entire system -- the simplest such observable is the symmetry operator $g$.


Notice that the conclusion of local indistinguishability does not even need the SW-SSB. What does need the SW-SSB  property is the local detectability of the symmetry breaking. Then we consider the convex sum of $\rho$ and $\tilde\rho$, 
\begin{equation}
\rho_+\equiv\frac{\rho+\tilde{\rho}}{2}.
\end{equation} 
The state explicitly breaks the strong symmetry down to a weak symmetry. 

The symmetry breaking does become {\color{black}locally} detectable using the fidelity average of a local charged operator if the original state $\rho$ has SW-SSB:
{
\color{black}
\begin{align}
&F\left(\rho_+, O^\dag(y)\rho_+ O(y)\right)\nonumber\\
&\geq\frac{1}{L^d}\sum_xF\left(\rho,O(x) O^\dag(y) \rho O(y) O^\dag(x)\right)\sim O(1),
\end{align}
where the inequality is ensured by the joint concavity of the fidelity \eqref{Eq: joint concavity}. It is straightforward to show that the above result on local detectability of explicit strong symmetry breaking for $\rho_+$ is also valid in the R\'enyi-$2$ sense. }

\section{Examples}
\label{sec:example}
We discuss a few examples in this Section. First in Sec.~\ref{sec:thermal} we conjecture, with explicit calculations in simple cases and plausibility arguments in general, that a thermal state within a fixed charge sector must have spontaneously broken strong symmetry, for any nonzero temperature. We then discuss two non-thermal examples of SW-SSB states in $(2+1)d$ and associated phase transitions: the $(2+1)d$ quantum Ising model under symmetric noise in Sec.~\ref{sec:Ising}, and a similar model with rotor degrees of freedom and $\U$ symmetry in Sec.~\ref{sec:rotor}. 

\subsection{SW-SSB in thermal states}
\label{sec:thermal}
A standard fact in statistical mechanics is the local equivalence between canonical and grand canonical ensembles. From the open system perspective, canonical ensembles by definition have strong symmetry, while grand canonical ensembles only have weak symmetry. The local equivalence suggests that the canonical ensembles have SW-SSB if the strong symmetry is not spontaneously broken in the conventional sense (i.e., the two-point function of a certain charged operator is long-range ordered).

{
\conjecture Given a local symmetric Hamiltonian $H$ and $0<\beta<\infty$, {if the Gibbs state $\rho_{\beta, \lambda} \propto P_\lambda e^{-\beta H}$ within a fixed charge sector described by a projector $P_\lambda$} has no SSB, then it must have SW-SSB. In particular, for sufficiently high temperatures the Gibbs state always has SW-SSB. 
}

Here a Gibbs state within a fixed charge sector is constructed by symmetrizing an ordinary Gibbs state. For example, when the strong symmetry $G$ is a finite group, and $\lambda: G\to\U$ is a one-dimensional representation, we express it as
\begin{equation}
    \rho = \frac{P_\lambda e^{-\beta H}}{\Tr\left[P_\lambda e^{-\beta H}\right]},
\end{equation}
where $P_\lambda = \frac{1}{|G|} \sum_{g \in G} {\lambda^{-1}(g)} g$ is the projector into the charge sector (one-dimensional rep.) $\lambda : G \to \U$.

We demonstrate the validity of the above conjecture in a simple commuting projector Hamiltonian of Ising spins in a canonical ensemble. On a lattice, the  Hamiltonian takes the following form:
\begin{align}
H_{\mathrm{CP}}=\sum\limits_{i}B_i
\end{align}
where $B_i$ satisfies 
\begin{equation}
    B_i^2=1, ~ [B_i, B_j]=0.
\end{equation}
We further assume $B_i$ is finitely supported around site $i$. The strong $\Z_2$ symmetry operator is $U=\prod_jX_j$ and the corresponding charged operator is $Z_i$, without loss of generality. We assume that 
\begin{equation}
  [B_i, U]=0,  ~ Z_i B_j = (-1)^{\delta_{ij}} B_j Z_i.
\end{equation}
For example, we can have $B_i=X_i$, or $X_i$ times any $\Z_2$-invariant functions of $Z_i$. We also assume that $B_i$ is a complete set of (mutually commuting) observables. 
 
Then one of the corresponding Gibbs canonical ensemble at finite temperature can be formulated as:
\begin{align}
\rho=\frac{1}{Z}\frac{1+U}{2}e^{-\beta H_{\mathrm{CP}}}.
\end{align}

If we choose the $B_i$-eigenbasis, the density matrix can be reformulated as
\begin{align}
\rho=\frac{1}{Z}\sum'\limits_{\{b_i\}}e^{-\beta\sum\limits_ib_i}|\{b_i\}\rangle\langle\{b_i\}|,
\end{align}
where $b_i=\pm1$ labels the eigenvalue of $B_i$, and the partition function $Z=\sum'_{\{b_i\}}e^{-\beta\sum_ib_i}$. Here the prime means that the sum is over configurations satisfying $\prod_i b_i=1$. Then the fidelity can be calculated easily.  For any $i\neq j$ we find: 
\begin{align}
\label{eq:FIsingbeta}
F(i,j)&=\frac{1}{\cosh^2 \beta}(1+\tanh^{N} \beta)^{-1}\nonumber\\
&\stackrel{N\rightarrow\infty}{\approx} \frac{1}{\cosh^2\beta}.
\end{align}
 Here we have taken the thermodynamic limit $N\rightarrow \infty$. We see that at zero temperature ($\beta\rightarrow\infty$), the fidelity $F(i,j)$ vanishes, which is consistent with the ground state of a commuting projector Hamiltonian having no SSB. But for finite temperature, the fidelity $F(i,j)$ will always be positive and independent of the distance between $i$ and $j$. On the other hand, a $\Z_2$ charged operator, such as $Z_i$, must anti-commute with at least one of $B_i$. This results in a correlation function $\Tr(\rho Z_i Z_j) = 0$ that vanishes at large $|i-j|$. Therefore, we conclude that a Gibbs canonical ensemble of the Hamiltonian $H=-\sum_i B_i$ at any finite temperature has SW-SSB of the $\Z_2$ symmetry.

We now provide a plausibility argument supporting Conjecture~$1$ for a generic local Hamiltonian $H$ and its Gibbs state $\rho$ in the canonical ensemble. We examine the replicated correlation function
\begin{widetext}
\begin{align}
    &F^{(n)}\equiv\frac{\Tr [(\sqrt{\rho}O(x)O^{\dagger}(y)\rho O(y)O^{\dagger}(x)\sqrt{\rho})^{n}]}{\Tr \rho^{2n}} = \frac{1}{Z(2n\beta)}\Tr [(O(x)O^{\dagger}(y)e^{-\beta H}O(y)O^{\dagger}(x)e^{-\beta H})^n ] \\
    &= \langle [O(x,\tau=0)O^{\dagger}(y,\tau=0)][O^{\dagger}(x,\tau=\beta)O(y,\tau=\beta)] [O(x,\tau=2\beta)O^{\dagger}(y,\tau=2\beta)] \cdots\rangle_{2n\beta}\nonumber,
\end{align}
\end{widetext}
where $A(\tau) \equiv e^{- \tau H} A e^{\tau H}$ and $\langle \ldots \rangle_{2n\beta}$ means the thermal (imaginary time) expectation value at inverse temperature $2n\beta$. Assuming no off-diagonal long-range order at temperature $2n\beta$ (otherwise even the weak symmetry is spontaneously broken), the above correlation function should factorize for large $|x-y|$:
\begin{align}
F^{(n)}&= \langle O(x,\tau=0)O^{\dagger}(x,\tau=\beta)O(x,\tau=2\beta) \cdots \rangle_{2n\beta}  \nonumber\\
& \cdot \langle O^{\dagger}(y,\tau=0)O(y,\tau=\beta)O^{\dagger}(y,\tau=2\beta) \cdots \rangle_{2n\beta},
\end{align}
which in general should be nonzero ($\sim O(1)$) unless fine-tuned. In particular, for low temperature or large $\beta$ (but still assuming no off-diagonal long-range order), the correlator should decay exponentially in imaginary time, so we expect
\begin{equation}
\label{eq:FnLargebeta}
    F^{(n)}\sim e^{-2n\beta\Delta},
\end{equation}
where $\Delta$ is the energy gap of the excitation created by $O(x)$ above the ground state. As long as $\beta\neq\infty$, the above expression is nonzero.

Analytically continuing to $n=1/2$, $F^{(n)}$ becomes the fidelity correlator, which demonstrates our original proposal. In particular, Eq.~\eqref{eq:FnLargebeta} agrees with Eq.~\eqref{eq:FIsingbeta} with $\Delta=2$ being the excitation gap.

Another interesting feature of Eq.~\eqref{eq:FnLargebeta} is that the SW-SSB disappears below temperature scale $T^*\sim \Delta/L$. In particular, this means that as far as SW-SSB is concerned, the zero-temperature limit $T\to 0^+$ and the thermodynamic limit $L\to\infty$ do not commute. This non-commutativity is absent in ordinary observables, at least when the ground state is gapped $\Delta\sim O(1)$  -- recall that in the imaginary-time path integral, these two limits are simply infinite size limits in different directions.

\subsection{$\Z_2$ SW-SSB in the quantum Ising model}
\label{sec:Ising}
Consider a (2+1)$d$ qubit system on a square lattice, with a strong $\Z_2$ symmetry $U=\prod_iX_i$. 
The system is initialized in a pure product state $|\psi_0\rangle=\otimes_i|X=+1\rangle_i$. Then we consider a nearest-neighbor dephasing channel that preserves the strong $\Z_2$ symmetry, $\E=\prod_{\langle ij\rangle}\E_{ij}$, with
\begin{align}
\E_{ij}[\rho]=(1-p)\rho+pZ_iZ_j\rho Z_iZ_j.
\label{ZZ channel}
\end{align}
{We note that this model is closely related to the $\Z_2$ toric code under bit-flip noise~\cite{TQM, fan2024diagnostics}. In fact, under the Kramers-Wannier duality mapping, they are exactly mapped to each other.}

To detect the possible SW-SSB and identify the universality class of the symmetry-breaking transition, we consider the fidelity correlator of $Z$, which is charged under the strong $\Z_2$ symmetry. 

{
\color{black}
The starting point is the decohered density matrix in the $\Z_2$ charge basis:
\begin{equation}
    \rho=\sum_{\braket{X}}Z(X)\ket{X}\bra{X}.
\end{equation}
Here $X$ denotes a configuration of $\Z_2$ charges (i.e. $X=\pm 1$), subject to the constraint that $\prod_i X_i=1$. In the following we write $X_i=(-1)^{n_i}$ with $n_i=0,1$. To compute the weight $Z(X)$, define an edge $\Z_2$ variable $s_{ij}=0,1$, which keeps track of the Kraus operator on the edge $ij$. When applied to the initial state $\ket{X_i=1}$, the resulting charge at site $i$ is given by
\begin{equation}
    (\nabla\cdot s)_i = \sum_{\mu=x,y}s_{i,i+\mb{e}_\mu}+s_{i-\mb{e}_\mu,i} \text{ mod }2.
\end{equation}
Here $\mb{e}_\mu$ is the unit vector along the $\mu$ direction.
We obtain
\begin{align}
    Z(X)&=\sum_{\nabla\cdot s=n} \prod_{\langle ij\rangle}(1-p)^{1-s_{ij}}p^{s_{ij}} \nonumber \\
    &=(1-p)^{N_e}\sum_{\nabla\cdot s=n} \Big(\frac{p}{1-p}\Big)^{\sum_{\langle ij\rangle}s_{ij}}.
\end{align}
To resolve the constraint $\nabla\cdot s=n$, we denote a special solution of this equation by $s^0$, and $s_{e}=s^0_e+\tilde{s}_p+\tilde{s}_q$ mod 2 (where $e$ stands for a nearest-neighbor edge, and $p,q$ are the two adjacent square plaquettes). $\tilde{s}_p$ are dual variables living on the plaquettes $p$. It is more convenient to work with $\mu_e=(-1)^{s_e}$ and $\tau_p=(-1)^{\tilde{s}_p}$, so $\mu_e=\mu_e(X) \tau_p\tau_q$, with $\mu_e(X)=(-1)^{s_e^0}$. Therefore
\begin{equation}
    Z(X)=[p(1-p)]^{N_e/2}\sum_{\tau}e^{\beta \sum_{\langle pq\rangle}\mu_{pq}(X) \tau_p\tau_q}
\end{equation}
where $e^{-2\beta}=\frac{p}{1-p}$. In other words, $Z(X)$ is the partition function of a 2D classical Ising model coupled to background $\Z_2$ gauge fields $\{\mu(X)\}$, with fixed gauge-invariant $\Z_2$ fluxes given by $X$.

Since $\rho$ is already diagonalized in this basis, we can directly compute the fidelity correlator. We have
\begin{align}
    \sigma & = Z_i Z_j \rho Z_i Z_j \nonumber \\
    & = \sum_X Z(X\cdot\delta X)\ket{X}\bra{X}.
\end{align}
where $\delta X_k=-1$ if $k=i,j$ and 1 otherwise,  and the fidelity correlator is found to be~\cite{chen2024separability}
\begin{align}
    F_{ij}&=\Tr \sqrt{\sqrt{\rho}\sigma\sqrt{\rho}}\\
    &=\sum_X \sqrt{Z(X) Z(X\cdot\delta X)} \nonumber \\
    &=\sum_X Z(X)\left(\frac{Z(X\cdot\delta X)}{Z(X)}\right)^{1/2} \nonumber \\
    &=\sum_X Z(X)e^{-\frac12\Delta F_{\mathrm dw}} \nonumber \\
    &=\braket{e^{-\frac12\Delta F_{\mathrm dw}}}_{\rm RBIM}\geq e^{-\frac12 \braket{\Delta F_{\rm dw}}_{\rm RBIM}}.
\end{align}
Here $\Delta F_{\rm dw}=-\ln \frac{Z(X\cdot\delta X)}{Z(X)}$ is the free energy cost of inserting a domain wall of length $\sim |i-j|$. $\braket{\cdots}_{\rm RBIM}$ stands for disorder average in the random-bond Ising model along the Nishimori line~\cite{TQM}. For $p>p_c$, the RBIM is in the paramagnetic phase (i.e. high temperature), so the disorder-averaged free energy cost for a domain wall approaches a constant. Thus we have shown that the fidelity correlator is long-range ordered for $p>p_c$. It also suggests that $F_{ij}$ decays exponentially for $p<p_c$, since the free energy cost diverges linearly with the distance in the ferromagnetic (low-temperature) phase. 

We note that the same results follow for the other definition of SSB in terms of sandwiched R\'enyi entropy. offering supporting evidence for our earlier conjecture in Sec.~\ref{Sec:fidelity} that SW-SSB defined using distinct distinguishability measures -- namely, quantum relative entropy and the fidelity correlator -- ought to be equivalent.
}

If we instead consider the R\'enyi-2 correlator \eqref{eq:R2}, following a similar derivation the correlation function can be mapped to the statistical mechanics model originating from evaluating the purity of the density matrix, $\Tr(\rho^2)$, which is exactly the partition function of 2$d$ classical Ising model. Therefore, the transition between $\Z_2$ strongly symmetric state and an SW-SSB state induced by a symmetric quantum channel will happen at $p_c^{(2)}\approx 0.178$. 

Furthermore, this is another example of the inequivalence between two different definitions of SW-SSB through fidelity and R\'enyi-2 correlators. For this model, the long-range R\'enyi-2 correlator is a stronger condition than the long-range fidelity correlator, since $p_c<p_c^{(2)}$. We note that it has been proposed in Ref.~\cite{Lee_2023} to use R\'enyi-$1$ quantity, such as the Fisher information or von Neumann relative entropy to define an ``intrinsic'' transition; based on its correspondence to the decodability transition in toric code \cite{TQM}, this transition was claimed to be the random-bond transition at $p_c$.


Our discussions are not limited to the Pauli channel like Eq. \eqref{ZZ channel}. For example, consider the following nearest-neighbor non-Pauli channel that is $\Z_2$ strongly symmetric,
\begin{align}
\E_{ij}(\rho)=(1-p)\rho+pe^{i\theta Z_iZ_j}\rho e^{-i\theta Z_iZ_j},
\label{theta channel}
\end{align}
and $\E=\prod_{\langle ij\rangle}\E_{ij}$. Notice that for $\theta=\pi/2$ we recover the previous channel \eqref{ZZ channel}. A more general channel similar to Eq. \eqref{theta channel} is
\begin{align}
    \mathcal{E}_{ij}(\rho) = \sum_{n} p_ne^{i\theta_n Z_i Z_j} \rho e^{-i\theta_n Z_i Z_j}, 
    \label{eq:general Ising channel}
\end{align}
where $p_n\geq 0$ and  $\sum_n p_n=1$.

For these non-Pauli channels, the density matrix is no longer diagonal in the $X$ basis. In the following, we will argue that nevertheless the Ising SW-SSB transition is still described by the RBIM universality class, the same as the Pauli case. We will utilize the replica trick to calculate the fidelity, which is not rigorous but still suggestive. In Appendix \ref{app:replica-free} we perform the calculations in the Pauli case $\theta=\pi/2$, and reproduce the same result as the exact calculation above, providing evidence for the validity of the replica trick. 

We define the replicated fidelity as
\begin{align}
F_{m,n}(x,y)=\Tr[(\rho^m\sigma\rho^m)^n],
\label{eq:replicatedfidelity}
\end{align}
and the standard definition of fidelity \eqref{eq:sqrtR2} is recovered by taking the limit $m,n\rightarrow 1/2$. Thus to compute $F_{m,n}(x,y)$, we 
need to study the replicated partition function $\Tr \rho^t$. It is not difficult to see that $t=(2m+1)n$, so the limit $m,n\rightarrow 1/2$ corresponds to $t\rightarrow 1$. 

Let us consider the domain wall basis (i.e. eigenbasis of $Z$), in which it is easier to compute the matrix elements of $\rho$. The initial pure state can be represented in the domain wall basis as $\rho_0 \propto \sum_{ D, D' } | D \rangle \langle D' |$, where the summation is taken over all possible $\Z_2$ domain wall configurations.  The decohered density matrix is 
\begin{equation}
\rho \propto \sum_{ 
D,D' } | D \rangle \langle D' | f(D,D'),
\end{equation}
Here we need to explain $f(D, D')$. The domain wall configuration can be labeled by $s_e=\pm 1$ on nearest-neighbor lattice edges $e$ ($s_e=1$ means no domain wall, $s_e=-1$ means a domain wall crossing the bond). Then an explicit calculation gives 
\begin{equation}
    f(D,D')=\prod_e \alpha^{\frac{s_e-s_e'}{2}} \beta^{\frac{|s_e-s_e'|}{2}}.
\end{equation}
Here 
\begin{equation}
\begin{gathered}
\beta= |1-p+pe^{2i\theta}| = \sqrt{1-4p(1-p)\sin^2\theta}, \\
\alpha=\frac{1-p+pe^{2i\theta}}{\beta}
\end{gathered}.
\end{equation}
Note that $\beta>0$ is real, and $\alpha$ is a pure phase factor. For a special point $\theta=\pi/2$, we have $\beta=1-2p, \alpha=1$.
In this case, $\beta^{\frac{|s_e-s_e'|}{2}}=\beta^{\frac{1-s_es_e'}{2}}$, and $\prod_e \beta^{\frac{1-s_es_e'}{2}} =  \beta^{|D-D'|}$, where $|D-D'|$ is the length of the relative domain wall, namely $|D-D'|=\sum_e \frac{1-s_es_e'}{2}$. 

Naively, the $\alpha^{\frac{s_e-s_e'}{2}}$ part complicates things, and the weight can not be expressed in terms of relative domain walls alone. However, if we consider the replicated density matrix:
\begin{align}
\Tr \rho^t &\propto\sum_{\{  D_s \}} f(D_1,D_2)f(D_2, D_3)\cdots f(D_{s},D_1) \nonumber\\
&= \prod_e \alpha^{\sum_k\frac{1}{2}(s_e^{(k)}-s_e^{(k+1)})} \prod_e \beta^{\sum_k\frac{1}{2}\big|s_e^{(k)}-s_e^{(k+1)}\big|}\nonumber\\
&= \prod_e \beta^{\sum_k\frac{1}{2}\big|s_e^{(k)}-s_e^{(k+1)}\big|}.
\end{align}

Crucially, the replicated partition function takes the same form regardless of the value of $\theta$, just with different values of the ``tension" $\beta$.
 This is true for each $t\geq 2$, and if we analytically continue to $t\rightarrow 1$, it makes sense to conclude that the partition function in the limit $t\rightarrow 1$ should have the same universal behavior as the $\theta=\pi/2$ case. Thus if the replica limit exists, the SW-SSB transition in the non-Pauli model defined in Eq. \eqref{theta channel} is described by the same RIBM universality class.

This can be generalized to the channel in Eq. \eqref{eq:general Ising channel}, with
\begin{equation}
    \beta = |\sum_n p_n e^{2i\theta_n}|, \alpha=\frac{\sum_n p_n e^{2i\theta_n}}{\beta}.
\end{equation}
Then the same result follows. {These calculations suggest that the ``universality class" of the Ising SW-SSB transition is described by the RBIM, independent of microscopic details.}

\subsection{U(1) SW-SSB  in a quantum rotor model}
\label{sec:rotor}
We now consider the analogous transition in a U(1) symmetric model. On each site $i$, there is a quantum rotor degree of freedom with the $2\pi$-periodic phase variable $\theta_i$ and the conjugate number $n_i$. They satisfy the canonical commutation relation $[\theta_i, n_j]=i\delta_{ij}$. The U(1) symmetry is generated by $Q=\sum_i n_i$. Initially, the system is in a pure product state $\ket{\psi_0}=\otimes_i\ket{n_i=0}$. Then we consider a nearest-neighbor dephasing channel that preserves the strong $\U$ symmetry $\E=\prod_{\langle ij\rangle}\E_{ij}$, with
\begin{align}
\E_{ij}(\rho)=\sum_{k\in \mathbb{Z}}p_ke^{ik(\theta_i-\theta_j)}\rho e^{-ik(\theta_i-\theta_j)}.
\end{align}
Here $0\leq p_k\leq 1$ should satisfy $\sum_k p_k=1$. Physically, we expect that $p_k$ should decay rapidly with $|k|$, and it is natural to impose ``charge-conjugation" symmetry so that $p_{-k}=p_k$. To make the problem analytically tractable we choose $p_k \propto e^{-\alpha k^2}$, but we expect the results should apply to other $p_k$ as long as it decays sufficiently rapidly as $|k|\rightarrow \infty$. 

We will study the phase diagram as a function of $\alpha$. Heuristically, when $\alpha$ is large, the channel is close to the identity, so the system remains close to the pure state. As $\alpha$ decreases, one expects a transition to a U(1) SW-SSB phase characterized by the long-range fidelity correlator of the U(1) order parameter $e^{i\theta}$. 

It proves useful to work with dual link variables defined as
\begin{equation}
    n_i = (\nabla\cdot E)_i, ~\theta_i-\theta_j=A_{ij}.
\end{equation} 
Here $(\nabla\cdot E)_i$ is the lattice divergence: $(\nabla\cdot E)_i=\sum_{j}E_{ij}$. One can think of $A_{ij}$ as U(1) gauge fields and $E_{ij}$ as the electric fields.

Expanding the quantum channel, we find that the initial state $\rho_0$ (which has no charge) is acted by paths of $e^{\pm ikA}$ operators, i.e., Wilson lines, which will create some configuration of electric charges.  Introduce an integer variable $m_e$ on each edge $e$ (with its canonical orientation), corresponding to acting by the $e^{im_e A_e}$ Kraus operator. We can interpret $m$ as some kind of ``electric" field, with the ``error" charges given by $(\nabla\cdot m)_i$. Denote the Wilson operator by $W_A(\bs{m})=\prod_e e^{im_e A_e}$. 
Then the decohered density matrix becomes
\begin{equation}
\rho = \sum_{\bs{m}}  P(\bs{m}) W_A(\bs{m}) \rho_0 W_A^\dag(\bs{m}), ~P(\bs{m}) =\prod_e p_{m_e}.
\end{equation}
Now we consider $\tr \rho^t$:
\begin{align}
&\Tr\rho^t = \\
&\sum_{\bs{m}^{(s)}}\prod_{s=1}^t P(\bs{m}^{(s)}) \Tr \left[\prod_{s=1}^t W_A(\bs{m}^{(s)}) |\psi_0\rangle \langle\psi_0|W_A^\dag(\bs{m}^{(s)})\right]\nonumber.
\end{align}
We rewrite the trace as
\begin{equation}
\prod_{s=1}^t \braket{\psi_0|W_A^\dag(\bs{m}^{(s)}) W_A(\bs{m}^{(s+1)})|\psi_0}.
\end{equation}
and see that the trace is non-vanishing only if $W_A (\bs{m}^{(s+1)})$ and $W_A (\bs{m}^{(s)})$ create the same charges. In other words, $\nabla \cdot (\bs{m}^{(s)}-\bs{m}^{(s+1)})=0$. Therefore we may write 
\begin{equation}
    m_e^{(s+1)}=m_e^{(1)}+ v_e^{(s)},
\end{equation}
where $v_e^{(s)}$ satisfies $\nabla \cdot v^{(s)}=0$. The ``R\'enyi" partition function becomes
\begin{equation}
\Tr\rho^t = \sum_{\bs{m}^{(1)}}P(\bs{m}^{(1)})\sum_{\bs{v}^{(s)}}\prod_{s=1}^{t-1} P(\bs{m}^{(1)}+\bs{v}^{(s)}). 
\end{equation}

We can resolve the constraint $\nabla \cdot v=0$ in the following way (suppressing the replica index):
\begin{equation}
    \prod_i\delta((\nabla\cdot v)_i)=\int_\phi e^{-i\sum_{i,\hat{\mu}}\phi_i(v_{i,\hat{\mu}}-v_{i-\hat{\mu},\hat{\mu}})}.
\end{equation}
We have defined the short-hand notation
\begin{equation}
\int_\phi\equiv \int_0^{2\pi}\prod_i \frac{d\phi_i}{2\pi}.
\end{equation}
Thus the partition function becomes
\begin{equation}
    \int_\phi  \sum_{{v}^{(s)}}P(m_e+v_e^{(s)})e^{-i\sum_{i,\hat{\mu}}v^{(s)}_{i,\hat{\mu}}(\phi^{(s)}_{i}-\phi^{(s)}_{i+\hat{\mu}})}.
\end{equation}
Here we have defined $m_e\equiv m_e^{(1)}$. The summation over the replica index $s=1,2,\cdots, t-1$ is kept implicit. Applying Poisson resummation, we find that the partition function becomes
\begin{equation}
   \int_\phi\sum_{\bs{m}, \bs{k}^{(s)}}e^{-\alpha \sum_e m_e^2} e^{-\frac{1}{\alpha}H_{\rm eff}[\bs{m}]},
   \label{eq:U1 partition function}
\end{equation}
with the effective Hamiltonian 
\begin{equation}
    H_{\rm eff}[\bs{m}]=\frac14\sum_{i,\hat{\mu}}(\phi^{(s)}_{i}-\phi^{(s)}_{i+\hat{\mu}}-2\pi k^{(s)}_{i,\hat{\mu}}-2i\alpha m_{i,\hat{\mu}})^2.
\end{equation}
This can be viewed as the replica partition function of the classical Villain model with imaginary bond disorder given by the $2i\alpha m_{i,\hat{\mu}}$ term. The effective ``temperature" is given by $\alpha$.

The bond disorder exhibits two unusual features. First it is not a continuous variable, rather valued in $2i\alpha\Z$. Second, it is purely imaginary. We shall however consider two limiting cases. First, for large $\alpha$, the distribution of $m_e$ is sharply peaked at $m_e=0$, and the effective temperature is also high, so once setting $m_e=0$ the statistical mechanics model belongs to the high-temperature, ``disordered" phase of the Villain model, which agrees with our expectation that there is no SW-SSB when the channel is very close to the identity. Next, for a small $\alpha$, it is at least plausible that the physics is qualitatively approximated by treating $a_e=2\alpha m_e$ as a real variable,
 whose probability distribution is given by $e^{-\frac{1}{4\alpha}a_e^2}$. Since $\alpha$ is small, the disorder again becomes weak, and it is reasonable to expect that the statistical model should be in the ``superfluid" phase, i.e. SW-SSB, which should be robust to weak disorder at sufficiently low effective temperature. In two dimensions, the low-temperature phase is the Kosterlitz-Thouless (KT) phase, so in this case, the model provides an example of quasi-long-range SW-SSB. Therefore under these assumptions, we find a SW-SSB phase for small $\alpha$ and a strongly symmetric phase for large $\alpha$.

To further substantiate the result, we also study the R\'enyi-2 transition of this model.  We find that $\Tr \rho^2$ can be mapped to a variant of the Villain XY model. In fact, for small $\alpha$, the Boltzmann weight is well-approximated by that of the Villain XY model. We thus expect that there is a R\'enyi-2 $\U$ SW-SSB phase for sufficiently small $\alpha$, and a transition at a finite value of $\alpha$ (details can be found in Appendix~\ref{app:U(1)}).

Let us also compare the results with a rotor model at finite temperature. Following the general argument presented in Sec.~\ref{sec:thermal}, in a thermal state the fidelity correlator is long-range ordered, independent of dimensionality. However, the decohered rotor model can exhibit a power-law fidelity correlator in the SW-SSB phase in 2D, fundamentally distinct from the thermal state.

\section{Recoverability}
\label{sec:recover}

To define concrete phases of matter for mixed quantum states, we need to show that two mixed states belonging to the same phase are two-way connected by low-depth symmetric local quantum channels. In other words, a channel on a state should be ``recoverable" if the state remains in the same phase.

Now we explore the recoverability of a mixed state with SW-SSB. Specifically, we demonstrate in the Ising example that SW-SSB, as defined in terms of the R\'enyi-2 correlator, could be recovered. This means that there exists a short-depth channel, whose depth is on the order of $\log L$ with $L$ representing the system size, which can transform a state with a long-range R\'enyi-2 (or more general R\'enyi-$n$) order parameter into a state with an exponentially decaying R\'enyi-2 correlator. However, SW-SSB in the fidelity sense is not recoverable due to the stability theorem. 

For simplicity, in this section, we focus on a special class of low-depth channels termed circuit channels. A circuit quantum channel comprises layers of disjoint local channels, or ``gates," forming a circuit. Such a channel is considered short-depth if the depth of the circuit is constant or at most sub-linear in $L$. Moreover, the channel is deemed symmetric if each local gate is strongly symmetric.

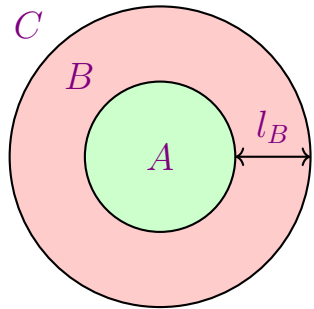
\begin{figure}
\begin{tikzpicture}[scale=1]
\tikzstyle{sergio}=[rectangle,draw=none]
\filldraw[fill=red!20, draw=black, thick] (0,0)circle (2);
\filldraw[fill=green!20, draw=black, thick] (0,0)circle (1);
\path (0,0) node [style=sergio,color=violet] {\Large$A$};
\path (1.5,0.4) node [style=sergio,color=violet] {\Large$l_B$};
\path (-1.075,1.075) node [style=sergio,color=violet] {\Large$B$};
\path (-1.75,1.75) node [style=sergio,color=violet] {\Large$C$};
\draw[thick, <->] (1,0) -- (2,0);
\end{tikzpicture}
\caption{The regions $A$, $B$, and $C$, with respect to which we calculate the CMI in Eq. \eqref{eq:CMIbound}.}
\label{Fig:recover}
\end{figure}

Recently, it has been highlighted in Ref. \cite{sang2024stability} that the effect of a local quantum channel $\E$ supported in a region $A$ as in Fig. \ref{Fig:recover} can be reversed using a local channel $\mathcal{R}_{\mathcal{E}}$ supported in an enlarged region $AB$, provided that the discrepancy in the conditional mutual information (CMI) between the initial state $\rho_i$ and the final state $\rho_f = \E(\rho_i)$ decays exponentially with respect to the width $l_B$ of the buffer region $B$, expressed as:
\begin{equation}
    I(A;C|B)_{\rho_i} - I(A;C|B)_{\rho_f} \simeq O(e^{-l_B/\xi}),
    \label{eq:CMIbound}
\end{equation}
where $\xi$ defines the correlation length of the mixed state. The local recovery channel $\mathcal{R}_{\mathcal{E}}$ is constructed through the Petz recovery channel. Two observations are important for our subsequent discussion: (1) The Petz recovery channel preserves the strong symmetry of the mixed state if the symmetry is on-site, which can be verified by the explicit form of the recovery channel (see Appendix \ref{app:recover}). 

(2) Building on this local recovery, a finite-depth circuit channel acting on the entire system can be recovered region by region, provided that at each intermediate step, the local gates being recovered can be recovered locally, \ie Eq. \eqref{eq:CMIbound} holds true for some chosen buffer region. The global recovery channel is also a circuit channel, with depth scaling logarithmically with system size $\log L$. We refer the readers to Appendix~\ref{app:recover} for details regarding the recovery channel.

\subsection{Recoveribility within the SSB phase}

We now revisit the Ising example in Sec.~\ref{sec:Ising}. We will establish that beginning from a pure symmetric product state, the two resulting mixed states, generated from symmetric dephasing in the form of Eq. \eqref{ZZ channel} with strengths $p_1$ and $p_2$, can be mutually connected by a short-depth circuit channel if $p_1$ and $p_2$ belong to the same phase of the RBIM. Assuming $p_1<p_2<\frac{1}{2}$, and denoting the two states as $\rho_{p_1}$ and $\rho_{p_2}$ respectively, it is evident that $\rho_{p_2}$ can be prepared from $\rho_{p_1}$ through an additional dephasing channel with strength
\begin{equation}
p = \frac{p_2-p_1}{1-2p_1}.
\end{equation}
Conversely, to construct a short-depth channel connecting $\rho_{p_2}$ to $\rho_{p_1}$, it is essential to demonstrate that this dephasing channel with strength $p$, when applied to $\rho_{p_1}$, can be recovered by a symmetric short-depth channel. This requirement is equivalent to ensuring that Eq. \eqref{eq:CMIbound} is met at every intermediate step of the recovery process.

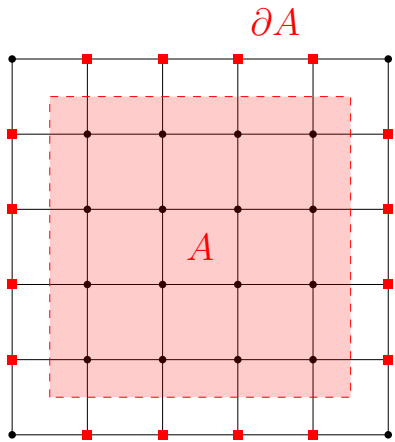
\begin{figure}
\begin{tikzpicture}[dot/.style={circle,fill=black,inner sep=0pt,minimum size=0.1cm}, 
        bdot/.style={rectangle,fill=red,inner sep=0pt,minimum size=0.12cm}]
    \foreach \i in {0,1,...,4} {
        \foreach \j in {0,1,...,5} {
            \draw[line width=0.1mm] (\i,\j) -- (\i+1,\j);
        }
    }
    \foreach \i in {0,1,...,5} {
        \foreach \j in {0,1,...,4} {
            \draw[line width=0.1mm] (\i,\j) -- (\i,\j+1);
        }
    }
    \foreach \i in {0,1,...,5} {
        \foreach \j in {0,1,...,5} {
            \node[dot] at (\i,\j) {};
        }
    }
    \foreach \i in {1,2,3,4} {
        \foreach \j in {0,5} {
            \node[bdot] at (\i,\j) {};
            \node[bdot] at (\j,\i) {};
        }
    }
    \draw[red,dashed,line width=0.1mm,fill=red,fill opacity=0.2] (0.5,0.5) rectangle (4.5,4.5);
    \path (2.5,2.5) node [color=red] {\Large$A$};
    \path (3.5,5.5) node [color=red] {\Large$\partial A$};
\end{tikzpicture}
\caption{The region $A$ for which we compute the von Neumann entropy encompasses all sites within the red dashed line. The boundary of $A$ includes all sites not within $A$ but adjacent to sites in $A$, marked with red squares above.}
\label{Fig:finiteregion}
\end{figure}

We now establish that within the same phase of the statistical model (\ie RBIM), Eq. \eqref{eq:CMIbound} holds. This assertion is supported by the following observation: The CMI, as a combination of von Neumann entropies from different regions, can be mapped to a linear combination of free energies of RBIM. Notably, the von Neumann entropy of a region $A$ (See Fig. \ref{Fig:finiteregion}) can be expressed as
\begin{equation}
S(A)_\rho = \lim_{t\to 1}\frac{1}{1-t} \ln \tr \rho_A^t,
\label{eq:freeenergy}
\end{equation}
where $\rho_A$ represents the reduced density matrix on $A$. From the expression of the decohered density matrix in Eq. \eqref{eq:ZZdecohered}, $\rho_A$ can be computed as
\begin{equation}
\rho_A = (1-p)^{N_A}\sum\limits_{\bs{l}_A}(\tanh\tau)^{|\bs{l}_A|}|\partial\bs{l}_A\rangle\langle\partial\bs{l}_A|,
\end{equation}
where $|\bs{l}_A|$ denotes all strings within the region $A$, potentially including links straddling between region $A$ and its complement. Let us define $Z_t^A = \tr (\rho_A^t)$ as the $t$-replica partition function, which can be expressed as
\begin{equation}
Z_t^A\propto\sum\limits_{\bs{l}_1,\cdots,\bs{l}_{t}}(\tanh\tau)^{|\bs{l}_1|+\cdots+|\bs{l}_t|}\langle\partial\bs{l}_1|\partial\bs{l}_2\rangle\cdots\langle\partial\bs{l}_{t}|\partial\bs{l}_1\rangle.
\end{equation}
The non-vanishing contributions to $Z_t^A$ arise from string configurations such that $\bs{l}_i$ and $\bs{l}_{i+1}$ differ only by closed loops \emph{relative to} the boundary of $A$. In other words, $\partial(\bs{l}_i-\bs{l}_{i+1})\cap A = \emptyset$. Following a derivation analogous to that leading to Eq. \eqref{Replica-t}, $Z_t^A$ represents the $t$-replica partition function of an RBIM (on the dual lattice) within region $A$ with a free boundary condition. In the limit $t\to 1$, the von Neumann entropy described in Eq. \eqref{eq:freeenergy} transforms into the free energy of an RBIM defined on region $A$ with a free boundary condition, denoted as $f_A$. Consequently, we can express the conditional mutual information as:
\begin{equation}
I(A;C|B) = f_{AB}+f_{BC}-f_{ABC}-f_B.
\label{eq:cmiasfreeenergy}
\end{equation}

\begin{figure}
\centering
\includegraphics[width=.48\textwidth]{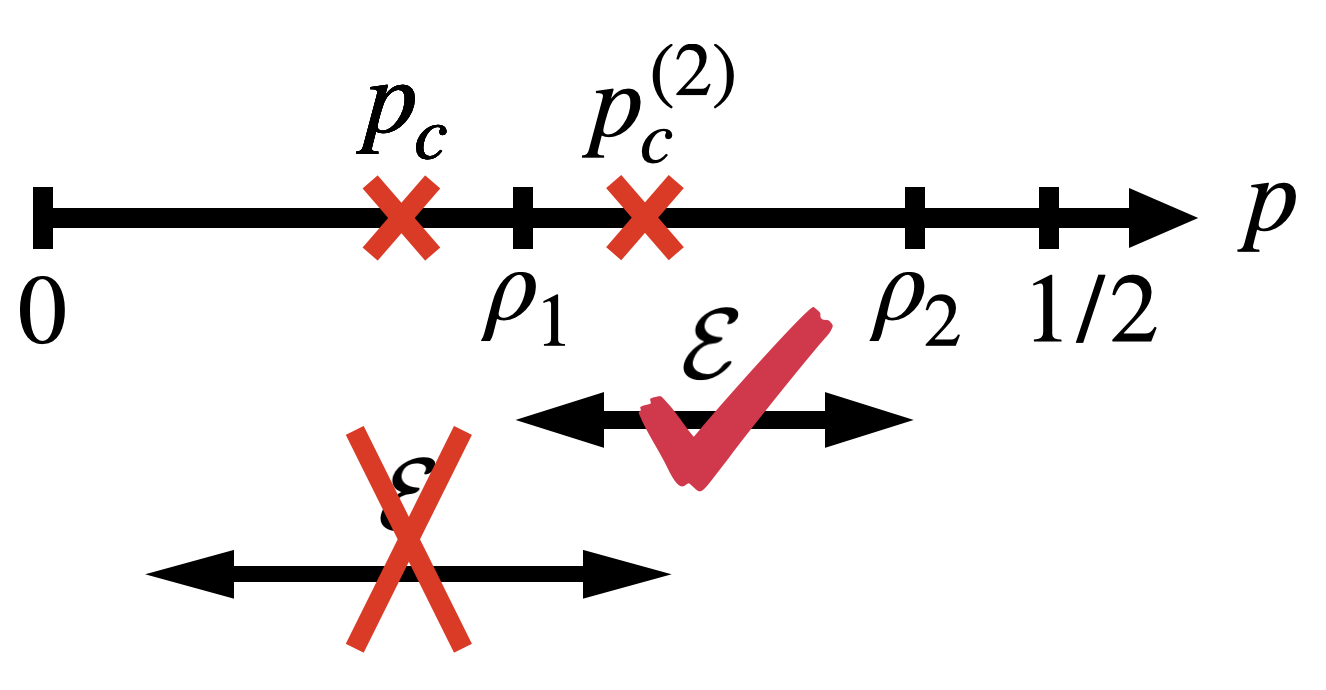}
\caption{Phase diagram of the decohered Ising model. Two states $\rho_1$ and $\rho_2$ in the same phase (both with $p>p_c$ in the figure) can be two-way connected through some low-depth channels. The two-way connectivity is obstructed if $p_1>p_c$ and $p_2<p_c$. The ``transition point'' for the R\'enyi-$2$ correlator, $p_c^{(2)}$, does not obstruct two-way connectivity.}
\label{fig:phasediagram}
\end{figure}

When attempting to recover the strength-$p$ dephasing channel that connects $\rho_{p_1}$ to $\rho_{p_2}$ region by region, at each intermediate step, the CMI can be mapped to the linear combination in Eq. \eqref{eq:cmiasfreeenergy}, computed in an RBIM with a spatially varying strength $p(x)$. Specifically, the regions recovered in previous steps have $p(x) = p_1$, while those that have not been recovered have $p(x) = p_2$. When $p_1$ and $p_2$ belong to the same phase of the RBIM, the system, despite its inhomogeneity, belongs to a definite phase (either ordered or disordered) and has a \emph{finite} correlation length. The free energy of a region $A$ thus scales as:
\begin{equation}
f_A = \int_{A} f[p(x)] d^2x + \int_{\partial A} g[p(x)] dl - \gamma_A,
\label{freeenergyscaling}
\end{equation}
where $f(p)$ is the free energy density of the RBIM with effective temperature $\mathrm{tanh}\beta = \frac{p}{1-p}$, and $g(p)$ denotes the boundary contribution. $\gamma_A$ is a constant depending on the phase of the statistical model.\footnote{In the ordered and disordered phases, $\gamma = \ln 2$ and $0$ when $A$ encompasses the entire system, respectively, and it vanishes when $A$ constitutes only a portion of the system.} Consequently, at each step, Eq. \eqref{eq:CMIbound} holds, where $\E$ represents the strength-$p$ dephasing within the region $A$ being recovered, and $\xi$ represents the correlation length of RBIM, with all terms in Eq. \eqref{freeenergyscaling} canceling out in Eq. \eqref{eq:cmiasfreeenergy}. Physically, the CMI, as a combination of free energies, should exhibit smooth variations within the same phase. Consequently, the LHS of Eq. \eqref{eq:CMIbound} is expected to decay exponentially, given that the system possesses a finite correlation length in both phases of RBIM. The key implication of our argument is that when $p_1$ and $p_2$ correspond to the same phase of the RBIM, $\rho_{p_1}$ and $\rho_{p_2}$ can be two-way connected by a symmetric channel within a depth of $\log L$. In particular, the critical decoherence strength for the R\'enyi-$2$ correlator, $p_c^{(2)}$, does not obstruct recoverability (Fig.~\ref{fig:phasediagram}).

\subsection{ Spontaneity and local irrecoverability} 
In the previous example, the CMI $I(A;C|B)$ in the SW-SSB phase is nonzero ($=\ln 2$) at large $l_B$. We now prove that such non-vanishing CMI is a universal feature of SW-SSB states, which is connected to the local non-recoverability of (symmetry-breaking) local channels.
\begin{theorem}
If a mixed state $\rho$ has SW-SSB in the fidelity correlator sense Eq.~\eqref{eq:sqrtR2}, then $\exists$ some local channel $\E_i$ (supported around site $i$) that cannot be recovered locally, namely $\nexists$ local $\E'_i$ such that $\E'_i\circ\E_i[\rho]=\rho$.

Furthermore, for any local $\E'_i$, the trace distance $D(\rho,\E'_i\circ\E_i[\rho])=O(1)>0$. As a consequence, the CMI $I(A;C|B)_{\rho}$ (in the geometry of Fig.~\ref{Fig:recover}) must remain nonzero at large distance between $A$ and $C$, for a region $A$ containing the support of $\mathcal{E}_i$ around site $i$.
\label{thm:nonrecoverable}
\end{theorem}
\begin{proof}
Let us normalize the order parameter $O_i$ such that the following operation is a valid quantum channel:
\begin{equation}
    \E_i[\rho]\equiv \frac{1}{2}\rho+\frac{1}{2}O_i\rho O_i^{\dagger}+\frac{1}{2}\sum_a K_{i,a}\rho K_{i,a}^{\dagger},
\end{equation}
where the local Kraus operators $\{K_{i,a}\}$ are chosen so that $O_i^{\dagger}O_i+\sum_aK_{i,a}^{\dagger}K_{i,a}=I$. For example, if $O_i$ is unitary, then we can choose $K_i=0$. In the Ising example $O=Z$ and the channel is nothing but a strong $Z$-measurement.

Now consider another arbitrary channel around site $i$, call it $\E'_i$, and define
\begin{equation}
    \tilde{\rho}\equiv \E'_i\circ \E_i[\rho].
\end{equation}
To see that $\tilde{\rho}\neq\rho$, it suffices to show that for some site $j$ far away from site $i$, $F(\tilde{\rho},O_j^{\dagger}\tilde{\rho}O_j)=O(1)>0$, namely $\tilde{\rho}$ does not even have the strong symmetry. This claim follows simply from the joint concavity and the data processing inequality of fidelity:
\begin{eqnarray}
\label{eq:notrecovered}
F(\tilde{\rho},O_j^{\dagger}\tilde{\rho}O_j) &\geq&  \frac{1}{2}F(\E'_i[\rho],\E'_i[O_iO_j^{\dagger}\rho O_jO_i^{\dagger}])  \nn
    &\geq& \frac{1}{2}F(\rho,O_iO_j^{\dagger}\rho O_jO_i^{\dagger}) \nn
    &=&O(1)>0.
\end{eqnarray}

Next we examine more carefully the trace distance $D(\rho,\tilde{\rho})$. Using the triangle inequality of trace distance, we have
\begin{align}\label{eq:recovertracedistance}
&D(\rho,\tilde{\rho})\geq D(\rho,O_j^{\dagger}\rho O_j)-D(\tilde{\rho},O_j^{\dagger}\rho O_j) \\
&\geq D(\rho,O_j^{\dagger}\rho O_j)-D(\tilde{\rho},O_j^{\dagger}\tilde{\rho} O_j)-D(O_j^{\dagger}\rho O_j,O_j^{\dagger}\tilde{\rho} O_j)\nonumber. 
\end{align}
Now we use several simple observations: 
\begin{enumerate}[(1)]
\item $D(\rho, O_j^{\dagger}\rho O_j)=1$ by strong symmetry; 
\item $D(\tilde{\rho}, O_j^{\dagger}\tilde{\rho} O_j)=1-c$ for some $c>0$ due to Eq.~\eqref{eq:notrecovered}; 
\item using the sub-multiplicative property of trace norm, we have
\end{enumerate}
\begin{equation}
    D(O_j^{\dagger}\rho O_j,O_j^{\dagger}\tilde{\rho} O_j)\leq \Vert O_j\Vert^2 D(\rho,\tilde{\rho}),
\end{equation}
where $\Vert O_j\Vert\sim O(1)$ is the operator norm of $O_j$.
Putting all these relations back into Eq.~\eqref{eq:recovertracedistance}, we obtain
\begin{equation}
    D(\rho,\tilde{\rho})\geq c/(1+\Vert O_j\Vert^2)>0.
\end{equation}
The above relation holds for any attempted recovery channel $\E'$. In particular, if we choose $\E'_i$ to be the Petz recovery map $\R_i$, then we have (see Eq.~\eqref{eq:localrecoverymap} in Appendix~\ref{app:recover})
\begin{equation}
    D(\rho,\R_i\circ\E_i[\rho])\leq \sqrt{\ln 2\cdot I(A;C|B)_{\rho}},
\end{equation}
which immediately implies that the CMI $I(A;C|B)_\rho$ must remain nonzero at large ${\rm dist}(A,C)$.
\end{proof}

{

We dub this theorem \textit{the spontaneity of SW-SSB}, for the following reason. Recall in pure state SSB, the cat (GHZ) state $\ket{\uparrow \cdots \uparrow}+\ket{\downarrow \cdots \downarrow}$ is sometimes called ``unphysical'' in the literature, which really means that the state is extremely unstable towards local decoherence: if the environment ``measures'' the $Z$-spin at site $i$, the state will immediately collapse to either $\ket{\uparrow \cdots \uparrow}$ or $\ket{\downarrow \cdots \downarrow}$. In a real system, even if the measurement strength (the probability for performing the measurement, $\delta P$) from the environment is small at each site, the probability for the entire system to remain in the cat state vanishes exponentially with the system size. Even though almost any state will go through some deformation under decoherence, the important feature of cat state is that the post-measurement state (a product state $|{\rm Prod}\rangle$) is very ``far away'' from the original state (a cat $|{\rm GHZ}\rangle$), in the sense that there is no local deformation (local channel) near the measurement location that takes $|{\rm Prod}\rangle$ to $|{\rm GHZ}\rangle$. We therefore dub such local irrecoverability of local measurements ``(weak) spontaneity''. We note that weak spontaneity holds even for weak-to-trivial SSB, as the above logic applies equality well to the weakly symmetric state $\ketbra{\uparrow \cdots \uparrow}{\uparrow \cdots \uparrow}+\ketbra{\downarrow \cdots \downarrow}{\downarrow \cdots \downarrow}$.

The above notion of spontaneity is ``weak'' in the sense that the irrecoverability holds for certain measurement results. If we average over (or simply forget) the measurement results, the post-measurement state may become locally recoverable. For example, take the weak-to-trivial example $\rho=\ketbra{\uparrow \cdots \uparrow}{\uparrow \cdots \uparrow}+\ketbra{\downarrow \cdots \downarrow}{\downarrow \cdots \downarrow}$, the post-measurement density matrix is $\E[\rho]=(1/2)\rho+(1/2)Z_i\rho Z_i=\rho$. If the post-measurement density matrix (without post-selecting the measurement outcome) cannot be locally recovered to the original state, we dub the spontaneity ``strong'' -- this is a stronger statement since if the post-measurement state is locally recoverable for all measurement results, it is simple to construct a recovery channel that recovers the whole density matrix, by taking the convex sum of the corresponding recovery channel for each measurement result.

In this language, what we have shown in Thm.~\ref{thm:nonrecoverable} is that the SSB of strong symmetry (either to weak or to trivial) is \textit{strongly spontaneous}. Let us consider a pedagogical example of SW-SSB: $\rho_0\propto\mathbbm{1}+X$ as the SW-SSB of $\Z_2$ Ising symmetry (Sec. \ref{Sec:fidelity}). We perform a strong $Z$-measurement on a single qubit and recognize that $\rho_0$ will immediately collapse to the maximally mixed state $\mathbbm{1}$ that no longer has the strong Ising symmetry. Heuristically, the \textit{only} information encoded in the SW-SSB density matrix is the global charge sector fixing, i.e., $\rho_0$ is the maximally mixed state within the even charge sector of the Ising symmetry. Once we perform a strong $Z$-measurement, the information on global charge sector fixing is completely destroyed, and we can never restore this information with any local quantum channel. 


Furthermore, the spontaneity theorem also implies a non-vanishing CMI is a universal feature of the SW-SSB density matrix. Similar to the stability theorem, this result only requires the fidelity correlator to be nonvanishing, and is insensitive to the ordinary linear correlation function. This means that Thm.~\ref{thm:nonrecoverable} also holds for strong-to-trivial SSB -- the most familiar example is again the GHZ state, which has CMI $\ln 2$. Also, note that weak-to-trivial SSB does not have a similar property. For example, $\rho=\frac{1}{2}(\ketbra{\uparrow\cdots\uparrow}{\uparrow\cdots\uparrow}+\ketbra{\downarrow\cdots\downarrow}{\downarrow\cdots\downarrow})$ has vanishing CMI -- from earlier discussion, this is because weak-to-trivial SSB is only ``weakly spontaneous''.
}

For pure states, the spontaneity of SSB is quantitatively described by a divergent dynamical susceptibility. For example, the magnetic susceptibility of the GHZ state is divergent with respect to the external Zeeman field along the $Z$ direction. {\color{black}Following the initial version of our manuscript, }in Ref. \cite{zhang2024fluctuation} the authors proposed a  \textit{fidelity susceptibility} which diverges in the SW-SSB phase and remains finite in the symmetric phase.

\section{Summary and Outlook}
\label{Sec: summary}
In this work, we systematically discussed a new phenomenon in mixed quantum states, namely strong-to-weak spontaneous symmetry breaking (SW-SSB). Our key results are:
\begin{enumerate}[1.]
\item We defined SW-SSB through the long-ranged fidelity correlator Eq.~\eqref{eq:sqrtR2} of charged local operators. We proved (Thm.~\ref{thm:SW-SSB_stability} and \ref{thm:SW-SSB_stability2}) that SW-SSB is stable against strongly symmetric low-depth local quantum channels. These stability theorems established SW-SSB as a universal property of mixed-state quantum phases.

\item We also considered an alternative definition of SW-SSB, which appeared in previous literature \cite{Lee_2023, Ma:2023rji, Ma:2024kma}, by the long-ranged R\'enyi-$2$ correlator Eq.~\eqref{eq:R2} of charged local operators. The two definitions of SW-SSB are inequivalent: there are examples with long-range fidelity correlators but short-range R\'enyi-2 correlators, {  and vice-versa (Appendix \ref{app:counterexample})}. Moreover, there is no analogue of the stability theorems (Thm.~\ref{thm:SW-SSB_stability} and \ref{thm:SW-SSB_stability2}) for the R\'enyi-$2$ correlator, as we showed through an explicit example in Sec.~\ref{sec:recover}. So the SW-SSB defined through the R\'enyi-$2$ correlator is not a universal property of mixed-state quantum phases. However, a weaker statement holds: a state with SW-SSB in the R\'enyi-$2$ sense must still be nontrivial, in the sense that it is not symmetrically invertible (Thm.~\ref{thm:noninvertible}).

\item We proposed that SW-SSB is a generic feature of thermal states at nonzero temperature (Sec.~\ref{sec:thermal}). In particular, we conjectured that if the canonical ensemble (i.e., with fixed charge sector) of a local symmetric Hamiltonian without a strong-to-trivial SSB, then it must have SW-SSB. We demonstrated the statement in certain commuting-projector models and provided plausibility arguments for general Hamiltonians.

\item We also studied non-thermal examples of SW-SSB, in models where a pure symmetric state is subject to a strongly symmetric finite-depth channel. In particular, in Sec.~\ref{sec:Ising} we considered a (2+1)$d$ quantum Ising model under a strongly symmetric nearest-neighbor $ZZ$ channel and found that the Ising SW-SSB transition (measured by the fidelity correlator) is described by the 2$d$ random-bond Ising model along the Nishimori line. The SW-SSB transition is the ungauged version (through Kramers-Wannier duality) of the decodability transition of the (2+1)$d$ toric code model under the bit-flip noise. In addition, we also considered the U(1) SW-SSB in a model of quantum rotors. By mapping to a Villain model with imaginary bond disorder we argued that the model exhibits U(1) SW-SSB for a sufficiently strong channel.

\item Following Ref.~\cite{sang2024stability}, we discussed the recoverability of mixed states under symmetric low-depth channels. Specifically, a necessary condition for the recoverability of a channel circuit through a low-depth symmetric channel is the Markov gap condition Eq.~\eqref{eq:CMIbound}. For the example of the decohered Ising model in Sec.~\ref{sec:Ising}, we showed that the Markov gap condition is violated only across the SW-SSB transition (measured by the fidelity correlator). As a consequence, if two states belong to the same side of the SW-SSB transition, they belong to the same phase, in the sense that they are two-way connected to each other through Log-depth symmetric channels. 

{
 
\item We demonstrated the spontaneity of SW-SSB and its relation to the local irrecoverability of local symmetry-breaking channels. Specifically, we proved that for a density matrix with SW-SSB, a local quantum channel can induce a global change that cannot be locally recovered (Thm.~\ref{thm:nonrecoverable}). This further implied that a non-vanishing CMI is a universal feature of SW-SSB. 
}

\end{enumerate}

We end this work with some open questions:
\begin{enumerate}[1.]
\item Strong-to-weak SSB of higher-form symmetry: A topologically ordered state can be understood as the spontaneous symmetry breaking of some higher-form symmetry \cite{Gaiotto_2015, Wen_2019}. Then what about the SW-SSB of higher-form symmetry? This question will be addressed in a forthcoming work \cite{Zhang_2024_higher}.

\item Ground state SSB  comes with many dynamical features, such as Anderson tower and Goldstone modes (if the symmetry is continuous). It is natural to ask whether SW-SSB has similar dynamical consequences, for example in Lindbladian dynamics. Two prior works \cite{OgunnaikeFeldmeierLee2023, Ma:2024kma} addressed this issue partially, but a general picture is still lacking at this point. {\color{black}In particular, a follow-up work \cite{huang2024hydro} showed that the Goldstone mode of U(1) SW-SSB can be identified as the  diffusive hydrodynamic mode, using Schwinger-Keldysh effective field theory.}

\item Conventional SSB features various topological defects, such as domain walls, vortices, and skyrmions. A natural avenue for future work is to consider whether SW-SSB can have similar defects and to explore the consequences of such defects.

\end{enumerate}

\acknowledgements{
We thank Yimu Bao, Tarun Grover, Timothy Hsieh, Andy Lucas, Zhu-Xi Luo, Max Metlitski, Shengqi Sang, Brian Swingle, Ruben Verresen, Xiao-Gang Wen, Yichen Xu, Carolyn Zhang, Zhehao Zhang, and Yijian Zou for inspiring discussions. RM is especially grateful to Shengqi Sang for many discussions and for explaining Ref. \cite{sang2024stability}. JHZ and MC are grateful to Hong-Hao Tu and Wei Tang for discussions and collaborations on a related project. RM was supported in part by Simons Investigator Award number 990660 and by the Simons Collaboration on Ultra-Quantum Matter, which is a grant from the Simons Foundation (No. 651440).  JHZ and ZB acknowledge support from NSF under award number DMR-2339319. RM, JHZ, ZB, and CW are supported in part by the grant NSF PHY-2309135 to the Kavli Institute for Theoretical Physics. LAL acknowledges support from the Natural Sciences and Engineering Research Council of Canada (NSERC) through Discovery Grants. Research at Perimeter Institute (LAL and CW) is supported in part by the Government of Canada through the Department of Innovation, Science and Industry Canada and by the Province of Ontario through the Ministry of Colleges and Universities. MC acknowledges support from NSF under award number DMR-1846109.

\vspace{0.2in}
\noindent{\it Note added -- } We would like to draw the reader’s attention to a related work \cite{YouSWSSB2024} on SW-SSB to appear in the same arXiv listing.

\appendix

{ 
\section{Counterexample for SW-SSB only in the R\'enyi-2 sense.}\label{app:counterexample}
}

Here, we are going to show $N$-qubit states $\rho_N$ that are strongly symmetric under $\mathbb{Z}_2$, exhibit SW-SSB in the sense of Rényi-2 correlator (Definition \ref{def:renyi_ssb}), but not in the fidelity sense (Definition \ref{def:swssb}). The strategy will be to construct ``spin glass'' ensembles $\rho_N \propto \sum_{m=1}^{M_N} p_{m} \ketbra{\psi_m}{\psi_m}$, $\lim_{N \to \infty} M_N = \infty$, such that $\lim_{|x-y| \to \infty} \lim_{N \to \infty} F_{O, N}(x, y) = 0$ for all order parameters $O$, but with its Rényi-2 correlator satisfying
\begin{align}
\label{eq:glasscorrelators}
    \lim_{|x-y| \to \infty} \lim_{N \to \infty} R^{(2)}_{O', N}(x,y) & = \frac{\sum_m p_m^2 |o'_m|^2}{\sum_m p_m^2} > 0,
\end{align}
for the particular order parameter $O' = Z$, with $\lim_{|x-y| \to \infty} \braket{\psi_m | Z_x Z_y | \psi_n} = o'_m \delta_{mn}$. Furthermore, we require $\sum_m p_m$ to diverge but $\sum_m p_m |o'_m|, \sum_m p_m^2 |o'_m|^2$ and $\sum_m p_m^2$ to converge. In that way, at least for $o'$, the fidelity correlator goes to zero, since
\begin{equation}
    \lim_{|x-y| \to \infty} \lim_{N \to \infty} F_{O', N}(x,y) = \frac{\sum_m p_m |o'_m|}{\sum_m p_m} = 0,
\end{equation}
and we have dropped the normalization condition $\sum_m p_m = 1$ for simplicity of calculation.

Let us consider a qubit chain with $2N$ spins, and define the strong $\Z_2$ symmetry to only act on even spins, meaning the symmetry is generated by $X_e:=\otimes_{n}X_{2n}$. We now consider all length-$N$ bit strings $m_1m_2\cdots m_N$, and view each bit string as the binary representation of an integer $m\in\{1,2,\ldots,2^N\}$. We can then construct the density matrix $\rho_N = \sum_{m=1}^{2^N} p_{m} \ketbra{\psi_m}{\psi_m}$, where $p_m=m^{-2/3}$ and
\begin{eqnarray}
    |\psi_m\rangle&=&\frac{\one+X_e}{\sqrt{2}}|m_1\tilde{m}_1m_2\tilde{m}_2...\rangle, \nn
    |\tilde{m}_i\rangle:&=&\sqrt{\frac{1}{2}+\frac{1}{m}}|m_i\rangle+\sqrt{\frac{1}{2}-\frac{1}{m}}X_i|m_i\rangle,
\end{eqnarray}
with $|m_i\rangle$ defined in the $Z$-basis. At large $m$, $|\tilde{m}_i\rangle\approx|+\rangle+[(-1)^{m_i}/m]|-\rangle$, so we should have $o'_m\sim1/m^2$. The auxiliary spins on the odd sites are introduced to make sure that different $|\psi_m\rangle$'s are orthogonal w.r.t. $O'$, allowing the simple forms of the correlators as in Eq.~\eqref{eq:glasscorrelators}.

Given that $p_m\sim m^{-2/3}$ and $o'_m\sim1/m^2$, then $\sum_m p_m |o'_m|, \sum_m p_m^2 |o'_m|^2$ and $\sum_m p_m^2$ all converge to an $O(1)$ value, while $\sum_mp_m\sim m_{\rm max}^{1/3}\sim 2^{N/3}$. We conclude that while the R\'enyi-2 remains finite, the fidelity correlator for $O'=Z$ vanishes exponentially in the thermodynamic limit. For other order parameters, the fidelity correlator still goes to zero, as $\rho_N$ converges to the trivial state $\rho_{\rm odd} \otimes \ketbra{+}{+}_{\rm even}$, $\rho_{\rm odd} \propto \sum_m p_m \ketbra{m}{m}$, in the fidelity (or trace) distance as $N \to \infty$: 
\begin{equation}
    F(\rho_N, \rho_{\rm odd} \otimes \ketbra{+}{+}_{\rm even}) \approx \frac{\sum_m p_m (1 - N/2m^2)}{\sum_m p_m} \to 1.
\end{equation}

It is clear from this construction, however, that this sequence of states is highly artificial. Thus, it remains a question of whether the same can happen for states arising from more natural contexts, such as thermal or decohered states.

{ 
\section{Alternative derivation of the fidelity correlator}
}
\label{app:replica-free}

We present an alternative derivation of the fidelity correlator in the decohered 2D Ising model Eq. \eqref{ZZ channel} using the replica trick. 

{
\color{black}
Then we utilize the replica trick to calculate the fidelity: the replicated fidelity is defined as
\begin{align}
F_{m,n}(x,y)=\Tr[(\rho^m\sigma\rho^m)^n],
\label{eq:replicatedfidelity}
\end{align}
and the standard definition of fidelity \eqref{eq:sqrtR2} is recovered by taking the limit $m,n\rightarrow 1/2$. In the ``open string'' basis, the density matrix $\E[\rho_0]=\rho$ is given by the following expression:
\begin{align}
\rho=(1-p)^{2N_v}\sum\limits_{\bs{l}}(\tanh\tau)^{|\bs{l}|}|\partial\bs{l}\rangle\langle\partial\bs{l}|,
\label{eq:ZZdecohered}
\end{align}
where $\bs{l}$ labels the configurations of ``open strings'', $\tanh\tau=\frac{p}{1-p}$, $|\partial\bs{l}\rangle=\prod_{v\in\partial\bs{l}}Z_v|++\cdots+\rangle$, and $N_v$ is the total number of vertices. Then our task is the calculation of the (replicated) fidelity Eq.~\eqref{eq:replicatedfidelity}. Define $t=(2m+1)n$. For simplicity, we focus on the special case with $n=1$ but the derivation can be easily generalized to $n>1$.  We find
\begin{widetext}
\begin{align}
\Tr(\rho^m\sigma\rho^m)&\propto\sum\limits_{\bs{l}_1,\cdots,\bs{l}_{t}}(\tanh\tau)^{|\bs{l}_1|+\cdots+|\bs{l}_t|}\langle\partial\bs{l}_1|\partial\bs{l}_2\rangle\cdots\langle\partial\bs{l}_m|Z_xZ_y|\partial\bs{l}_{m+1}\rangle\langle\partial\bs{l}_{m+1}|Z_xZ_y|\partial\bs{l}_{m+2}\rangle\cdots\langle\partial\bs{l}_{t}|\partial\bs{l}_1\rangle\nonumber\\
&=\sum\limits_{\{\sigma_v^{(1)},\cdots,\sigma_{v}^{(t)}\}=\pm1}\sigma_x^{(m)}\sigma_y^{(m)}\prod\limits_e\prod\limits_{j=1}^t\left(1+\tanh\tau\prod\limits_{v\in\partial e}\sigma_v^{(j)}\right)=\langle\sigma_x\sigma_y\rangle_{H_{\mathrm{eff}}},
\end{align}
\end{widetext}
here we utilized the following formula:
\begin{align}
\langle\partial\bs{l}_1|\partial\bs{l}_2\rangle=\frac{1}{2^{N_v}}\sum\limits_{s_v}\left(\prod\limits_{v_1\in\partial\bs{l}_1}s_{v_1}\right)\left(\prod\limits_{v_2\in\partial\bs{l}_2}s_{v_2}\right),
\end{align}
and defined $\sigma_{v_j}^{(j)}=s_{v_j}^{(j)}s_{v_j}^{(j+1)}$, which leads to the constraint $\prod_{j=1}^t\sigma_{v}^{(j)}=1$ for all sites $v$. The replicated fidelity has been expressed as the correlation function of the following Hamiltonian at the inverse temperature $\beta=\tau$:
\begin{align}
H_{\mathrm{eff}}=-\sum\limits_{\langle i,j\rangle}\left(\sum\limits_{k=1}^{t-1}\sigma_i^{(k)}\sigma_j^{(k)}+\prod\limits_{k=1}^{t-1}\sigma_i^{(k)}\sigma_j^{(k)}\right).
\label{effective Hamiltonian}
\end{align}
For $n>1$ the derivation is very similar, and we find the replicated fidelity is the following correlation function of the Hamiltonian \eqref{effective Hamiltonian},
\begin{align}
F_{m,n}(x,y)= \left\langle\prod\limits_{\alpha=1}^n\sigma_x^{\alpha}\sigma_y^{\alpha}\right\rangle_{H_{\mathrm{eff}}}.
\end{align}
The qualitative behavior of $F_{m,n}(\rho,\sigma)$ is essentially determined by the effective Hamiltonian \eqref{effective Hamiltonian}. This can also be understood from the replica partition function $\Tr\rho^t$:
\begin{align}
\label{Replica-t}
&\Tr(\rho^t)\propto\\
&\sum\limits_{\mathclap{\{\sigma_v^{(1)},\cdots,\sigma_{v}^{(t)}\}}}\ \exp\left\{-\tau\sum\limits_{\langle i,j\rangle}\left(\sum\limits_{k=1}^{t-1}\sigma_i^{(k)}\sigma_j^{(k)}+\prod\limits_{k=1}^{t-1}\sigma_i^{(k)}\sigma_j^{(k)}\right)\right\}\nonumber,
\end{align}
which is exactly the partition function of the effective Hamiltonian $H_{\mathrm{eff}}$. In particular, if we take the limit $m,n\rightarrow1/2$, the replicated fidelity \eqref{eq:replicatedfidelity} will be recovered back to Eq. \eqref{eq:sqrtR2}, and effective Hamiltonian \eqref{effective Hamiltonian} becomes the Hamiltonian of random-bond Ising model (RBIM) along the Nishimori line \cite{fan2024diagnostics}. Then the transition between a $\Z_2$ strongly symmetric state at $p < p_c$ and an SW-SSB state induced by a symmetric quantum channel at $p > p_c$ happens at $p_c\approx 0.109$. \add{We note that the same result for the fidelity correlator can be established directly without using the replica trick, which is explained in Appendix \ref{app:replica-free}. } {Our results are fully consistent with the decodability transition of the toric code \cite{TQM, WangPreskill2003, bao2023mixedstate, fan2024diagnostics, chen2024separability}, which can now be interpreted as the ``gauged'' version of the SW-SSB transition through Kramers-Wannier transform  -- recall that gauge symmetry is naturally strong since the total gauge charge is constrained to be trivial in the Hilbert space. }
}

\section{$\U$ R\'enyi-2 SW-SSB in the rotor model }
\label{app:U(1)}

In this section we study the R\'enyi-2 SW-SSB transition in the quantum rotor model. 

The initial state $\rho_0$ is given by 
\begin{equation}
    \rho_0\propto\int_{\theta,\theta'} \ket{\theta}\!\bra{\theta'}.
\end{equation}
It is easy to see that 
\begin{equation}
    \E(\rho_0)\propto\int_{\theta,\theta'}  \prod_{\langle ij\rangle}G_{ij}(\theta,\theta')\ket{\theta}\!\bra{\theta'},
\end{equation}
where the factor $G_{ij}(\theta,\theta')$ is defined as
\begin{equation}
    G_{ij}(\theta,\theta')=\sum_k p_ke^{ik(\theta_{i}-\theta_j-\theta_i'+\theta_j')}.
\end{equation}
Therefore
\begin{equation}
    \tr \rho^2 \propto  \int_{\theta,\theta'}  \prod_{\langle ij\rangle}G^2_{ij}(\theta,\theta').
\end{equation}

 Let us write 
\begin{equation}
    G_{ij}^2(\theta, \theta')=\sum_{n}f_{n} e^{in(\theta_i-\theta_j-\theta_i'+\theta_j')}.
\end{equation}
Up to an overall constant, it is easy to see that $\Tr\rho^2$ is proportional to the following partition function:
\begin{equation}
    Z=\int \mathcal{D}\theta \prod_{\langle ij\rangle} \sum_{n_{ij}} f_{n_{ij}} e^{in_{ij}(\theta_i-\theta_j)}.
\end{equation}
This can be viewed as a variant of the classical XY model.

For $p_k\propto e^{-\alpha k^2}$, after some algebra we obtain
\begin{equation}
f_n\propto
\begin{cases}
    e^{-\frac{\alpha}{2}n^2} \vartheta_3(e^{-2\alpha}) & n\text{ even}\\
    e^{-\frac{\alpha}{2}n^2} \vartheta_2(e^{-2\alpha}) & n\text{ odd}
\end{cases}.
\end{equation}
Here $\vartheta_2$ and $\vartheta_3$ are Jacobi theta functions, defined as
\begin{equation}
    \vartheta_2(q)=\sum_{k\in \Z}q^{(k+\frac12)^2}, \vartheta_3(q)=\sum_{k\in \Z}q^{k^2}.
\end{equation}
A useful fact is that $\vartheta_3(q)$ and $\vartheta_2(q)$ are exponentially close when $q$ is close to 1 (the difference between them is $\sim \sqrt{\frac{\pi}{1-q}}e^{-\frac{\pi^2}{1-q}}$) . For small $\alpha$, we have $e^{-2\alpha}\approx 1$, so $f_n \propto e^{-\frac{\alpha}{2}n^2}$. Using Poisson resummation we find:
\begin{multline}
    \sum_{n_{ij}\in \Z} e^{-\frac{\alpha}{2}n_{ij}^2}e^{in_{ij}(\theta_i-\theta_j)}  \\
    =\sqrt{\frac{2\pi}{\alpha}}\sum_{m_{ij}\in \Z} e^{-\frac{1}{2\alpha}(\theta_i-\theta_j+2\pi m_{ij})^2},
\end{multline}
which is the familiar Villain model at temperature ${\alpha}$.  The approximation becomes asymptotically exact as $\alpha\rightarrow 0$. Thus we can conclude that for sufficiently small $\alpha$ there is R\'enyi-2 SW-SSB of the $\U$ symmetry. 

In addition, in two dimensions the model exhibits R\'enyi-2 quasi-SW-SSB.
On a square lattice, numerically it is known that the KT point is $\alpha_c\approx 1.353$~\cite{XYMC}. For this range of $\alpha$, the approximation $f_n\propto e^{-\alpha n^2/2}$ remains reasonable, so we expect that our model is qualitatively similar to the Villain model, exhibiting a KT transition from the quasi-SW-SSB phase to the strongly symmetric phase.

\section{The recovery channel}
\label{app:recover}
In this Appendix, we provide the details of the recovery channel as outlined in Sec \ref{sec:recover} and originally proposed in Ref. \cite{sang2024stability}. In particular, we aim to construct a short-depth recovery channel for a finite-depth quantum channel that preserves the gap of a mixed state. Here the gap of a mixed state is defined as follows.

{\definition A mixed state $\rho$ has a Markov gap $\Delta$ if the conditional mutual information for the three regions depicted in Fig.\ref{Fig:recover} decays exponentially, i.e.
\begin{align}
&I(A;C|B)_\rho\leq 2 \ln(\min \{ d_A,d_C \}) \exp (-\Delta\cdot\mathrm{dist}(A,C)),   \nonumber\\
& =\const\cdot \min ( |A|,|C| )\exp (-\Delta\cdot\mathrm{dist}(A,C)),
\end{align}
where $\mathrm{dist}(A,C)$ is the closest distance between the region $A$ and $C$. $|A|$ and $|C|$ represent the number of sites within regions $A$ and $C$, respectively.

}

We observe that any mixed state has a gap $\Delta\geq 0$ due to the local dimension bound.

\subsection{Recovery of a local channel}
\label{sec:localrecoverymap}

The first result states that, for a gapped mixed state, the effect of a local channel $\mathcal{E}$ acting on region $A$ can be recovered by another local channel $\mathcal{R}$, supported on a region $A\cup B$ surrounding $A$, up to exponential decaying errors. The argument goes as follows. For any state $\rho$, non-negative operator $\sigma$ and quantum channel $\mathcal{E}$, we have the data-processing inequality:
\begin{equation}
    D(\rho \Vert \sigma) - D(\E(\rho)\Vert \E(\sigma))\geq -2\log_2 F(\rho,\R_{\sigma,\E}\circ\E(\rho)),
\end{equation}
where $F$ is the fidelity, and 
\begin{equation}
    \R_{\sigma,\E}(\cdot):=\int_t \beta_0(t) \R^{\frac{t}{2}}_{\sigma,\E}(\cdot)
\end{equation}
is the rotated Petz recovery map \cite{junge2018universal}. Here $\beta_0(t) = \frac{\pi}{2}(\cosh(\pi t)+1)^{-1}$ and 
\begin{equation}
    \R^{t}_{\sigma,\E}(\cdot) := \sigma^{\frac{1}{2}-it}\E^\dagger[\E(\sigma)^{-\frac{1}{2}+it}(\cdot)\E(\sigma)^{-\frac{1}{2}-it}]\sigma^{\frac{1}{2}+it}.
    \label{eq:rotatedPetz}
\end{equation}
Let us assume: (1) $\E$ acts nontrivially only in region $A$; (2) $\sigma = \rho_{AB}\otimes \rho_C$. We then have
\begin{align}
        -&2 \log_2 F(\rho,\R_{\sigma,\E}\circ\E(\rho))\leq  I(AB;C)_\rho - I(AB;C)_{\E(\rho)} \nonumber\\
        =& I(A;C|B)_\rho - I(A;C|B)_{\E(\rho)}.
    \label{eq:localerror}
\end{align}
Due to the positive-semidefinite nature of the CMI, as well as the data processing inequality for local channels (acting on the non-conditioning systems), we have
\begin{equation}
    0\leq I(A;C|B)_{\E(\rho)}\leq I(A;C|B)_\rho. 
\end{equation}
Therefore, the LHS of Eq.~(\ref{eq:localerror}) is upper bounded by a non-negative number exponentially small in $\mathrm{dist}(A, C)\sim l_B$. By the Fuchs–van de Graaf inequality, at large distances, the constraint on the fidelity can also be expressed as a constraint on the trace distance,
\begin{align}
    \Vert \rho - \R_{\sigma,\E}\circ\E(\rho) \Vert_1&\leq  \sqrt{4\ln 2\cdot I(A;C|B)_\rho} \nonumber\\
    &\simeq \const\cdot e^{-\Delta\cdot\mathrm{dist}(A,C)/2}.
    \label{eq:localrecoverymap}
\end{align}
Eq. \eqref{eq:localrecoverymap} implies that when the underlying state $\rho$ has a gap $\Delta>0$, the impact of a local channel can be recovered locally, with an error decaying exponentially in the width of the buffer region $B$.

\subsection{Symmetries}

Now let us show that when the channel $\E$ is strongly symmetric under a unitary on-site symmetry, the associated recovery channel $\R$ is also strongly symmetric. Using the explicit form of the rotated Petz recovery map in Eq \eqref{eq:rotatedPetz}, one has
\begin{equation}
    \R^t_{\sigma,\E}(\cdot)=\rho_{AB}^{-it+\frac{1}{2}}\E^\dagger[\E(\rho_{AB})^{it-\frac{1}{2}}(\cdot)\E(\rho_{AB})^{-it-\frac{1}{2}}]\rho_{AB}^{it+\frac{1}{2}}.
\end{equation}
If we pick an arbitrary Kraus representation of $\E$, \ie $\E(\rho) = \sum_i K_i^\dagger \rho K_i$, a corresponding Kraus representation of the recovery map is
\begin{equation}
    K_{\R,i}^\dagger = \rho_{AB}^{-it+\frac{1}{2}}K_i\E(\rho_{AB})^{it-\frac{1}{2}}.
\end{equation}
In order for the recovery channel to be strongly symmetric, we only need $\rho_{AB}$ to be \emph{weakly} symmetric under $U_{AB}$, which is the symmetry operation restricted to the region $AB$. This can be explicitly verified as follows:
\begin{equation}
\begin{split}
    U_{AB} \rho_{AB} U_{AB}^\dagger = & U_{AB} \tr_{\overline{A\cup B}}(\rho) U_{AB}^\dagger \\
    = & \tr_{ \overline{A\cup B}}(U\rho U^\dagger)=\rho_{AB},
\end{split}
\end{equation}
where $U$ is the generator of the global symmetry, and the last line follows from the fact that the state $\rho$ is strongly symmetric. Applying a similar argument, one can also demonstrate that $U_{AB'}\E(\rho_{AB})U_{AB'}^\dagger= \E(\rho_{AB})$, as long as the channel $\E$ (acting on $A$) is strongly symmetric and we enlarge $AB$ to $AB'$ to accomodate the depth of $\E$. As a summary, each Kraus operator $K_{\R,i}^\dagger$ of the recovery channel commutes with $U$, therefore the recovery map is strongly symmetric.

\subsection{Global recovery}
\label{sec:globalrecovery}
Now, we proceed to demonstrate that a symmetric finite-depth circuit channel $\E$, which acts on the entire system and preserves the gap of the underlying mixed state, can also be recovered by a short-depth channel. We assume that the maximal width of gates in $\E$ is $w$, and $\E$ has $l$ layers of gates. The depth of $\E$ is then defined to be $wl$. The global recovery needs two crucial assumptions: (a) After each layer of $\E$, the state, referred to as $\rho_k$ with $0\leq k \leq l$, has a finite gap lower bounded by $4\Delta$. (b) Another important requirement is, if we act the state $\rho_{k-1}$ with any subset of gates in $k$-th layer of $\E$, the resulting state remains gapped, also lower bounded by $4\Delta$. In the context of the RBIM, this assumption implies that in the intermediate states, where a part of the system has been recovered, the entire system, despite its inhomogeneity, remains in a definite phase of the RBIM with a finite correlation length.

The strategy of the global recovery is to recover $\E$ layer by layer. We cut the system into hypercubes of linear size $L_0$ -- thus we have $(L/L_0)^d$ hypercubes in total, where $d$ is the spatial dimension of the system. Here $L_0$ is assumed to be larger than $w$ and the correlation length $1/2\Delta$. We first reorganize each layer of $\E$ into $2^d$ new layers. In other words, we now perceive the channel $\E$ as a circuit with $h=2^dl$ layers, but the gates become ``sparse." Specifically, in each new layer, within a hypercube of size $2L_0$, only a hypercube of size $L_0$ contains nontrivial gates. Crucially, for each new layer, these nontrivial hypercubes are separated by a minimum distance of $L_0$. In the first recovery step, we reverse all gates in the last ($h$-th) layer of the new circuit. We label the hypercubes undergoing recovery as $C_x$ with $x \in {1, \dots, (L/2L_0)^d}$, the gates in the $h$-th layer within $C_x$ as $\E_x^h$, and the corresponding recovery channels as ${\R^h_x}$. The recovery map $\R^h_x$ is chosen to be the local recovery map mentioned in Sec \ref{sec:localrecoverymap}, confined to a hypercube $C_x'$ with a linear size of $2L_0$, and sharing the same center as the hypercube $x$. Note that these recovery channels act on disjoint hypercubes, and can therefore be implemented simultaneously in a single layer. 
Once more, we denote the state after each new layer of $\E$ as $\rho_k$, now with $0 \leq k \leq 2^dl$. We have:
\begin{align}
        &\left\Vert \prod_x \R^h_x(\rho_h) - \rho_{h-1} \right\Vert_1 \nonumber\\
        \leq & \sum_{y=1}^{(L/2L_0)^d} \left\Vert \prod_{z=1}^y \R^h_z\circ\E_z^h(\rho_{h-1}) - \prod_{z=1}^{y-1} \R^h_z\circ\E_z^h(\rho_{h-1}) \right\Vert_1 \nonumber\\
        \leq & \sum_{y=1}^{(L/2L_0)^d} \left\Vert  \R^h_y\circ\E_y^h(\rho_{h-1}) - \rho_{h-1} \right\Vert_1 \nonumber\\
        \leq & \const\cdot\left(\frac{L}{2L_0}\right)^d\cdot \mathrm{Poly}(L_0) \cdot e^{-\Delta\cdot L_0},
    \end{align}
where we have used the triangle inequality and the data processing inequality of the trace distance (note that recovery channels with distinct labels act in disjoint regions, and therefore commute).

Finally, we proceed to recover the channel $\E$ layer by layer. At each step of the recovery, we reverse gates in $(L/2L_0)^d$ hypercubes with a linear size of $L_0$. Denote the $k$-th layer of gates in $\E$ as $\E^k$, and the corresponding recovery channel by $\R^k$. We have
\begin{equation}
    \begin{split}
        &\left\Vert \prod_{k=1}^{h} \R^k\circ\E(\rho) - \rho \right\Vert_1 \\
        = & \left\Vert \sum_{k=1}^h [\prod_{m=1}^{k} \R^m\circ\E^m(\rho)-\prod_{m=1}^{k-1} \R^m\circ\E^m(\rho)] \right\Vert_1 \\
        \leq & \sum_{k=1}^{h} \left\Vert \prod_{m=1}^{k-1} \R^m[\R^k\circ\E^k(\rho_{k-1}) - \rho_{k-1})] \right\Vert_1 \\
        \leq & \sum_{k=1}^{h} \left\Vert  \R^k\circ\E^k(\rho_{k-1}) - \rho_{k-1} \right\Vert_1 \\
        \leq & \const\cdot\left(\frac{L}{2L_0}\right)^d l\cdot \mathrm{Poly}(L_0)\cdot e^{-\Delta\cdot L_0},
    \end{split}
\end{equation}
Here, $\rho = \rho_0$ represents the original state, and all layers are time-ordered. Specifically, all layers in $\E$ are time-ordered, followed by layers in the recovery map with anti-time ordering. In the derivation we have used (1) the triangle inequality; (2) the data processing inequality for the trace distance; (3) the assumption that $\rho_k$ has a nonzero gap for all $k$. This is precisely the assumptions (b) in the first paragraph of this section. In order to recover the error to $O(\frac{1}{\mathrm{Poly}(L)})$, we need $L_0\simeq \ln(L)/\Delta$. As each gate of the recovery channel has a width of order $L_0$, the total depth of the recovery map is thus $O(\ln L)$.

\subsection{An illuminating counterexample}

We now discuss an illustrative example, in which the global recovery scheme fails due to violation of the condition on CMI Eq.~\eqref{eq:CMIbound} when the channel is acting on parts of the system.

Consider a GHZ state in a quibit chain
\begin{equation}
|\psi_0\rangle=|\uparrow\uparrow\cdots\rangle+|\downarrow\downarrow\cdots\rangle.
\end{equation}
Starting from $\rho_0=|\psi_0\rangle\langle\psi_0|$, we can perform a strong $X_i$ measurement on each site, represented by the (depth-$1$) channel $\mathcal{E}=\prod_i\E_i$ where
\begin{equation}
    \E_i[\rho]=\frac{1}{2}\rho+\frac{1}{2}X_i\rho X_i.
\end{equation}
It is easy to verify that the final state is $\rho_f=\E[\rho_0]\propto I+\prod_i X_i$.

The effect of $\E$ on $\rho_0$ should not be recoverable through a low-depth channel, since the connected correlation function $\tr(\rho Z_xZ_y)-\tr(\rho Z_x)\tr(\rho Z_y)$ at large $|x-y|$ is nonzero for $\rho_0$ but zero for $\rho_f$. However, naively the CMI calculated in the geometry of Fig.~\ref{Fig:recover} is $\log 2$ for both $\rho_0$ and $\rho_f$, which satisfies the requirement from Eq.~\eqref{eq:CMIbound}. So what went wrong? It turns out that the CMI bound condition Eq.~\eqref{eq:CMIbound} is violated when only part of the system is being acted by the channel. 

Specifically, let us consider the ``intermediate'' state (below we have two special sites located at two far-separated points $i=1,x$, the region $A$ is defined as $1<A<x$, and the region $B$ is defined as $x<B\leq L$)
\begin{eqnarray}
    \rho'&=&\prod_{i\neq1,x}\E_i[\rho_0] \nn
    &=&\frac{1}{2^{L-1}}\sum_{X_A X_B}[|\uparrow X_A\uparrow X_B\rangle+X_AX_B|\downarrow X_A\downarrow X_B\rangle], \nn
\end{eqnarray}
where $[\psi]$ stands for $\ketbra{\psi}{\psi}$. Acting on $\rho'$ with $\E_1\cdot \E_x$ will produce the final state $\rho_f\propto I+\prod_i X_i$. But this step from $\rho'$ to $\rho_f$ is not locally recoverable since it eliminates a Bell pair between $i=1$ and $i=x$. In terms of CMI, it is easy to check that for $\rho'$:
\begin{equation}
    I(i=1;x\cup B|A)=2\log2,
\end{equation}
which violates the CMI bound Eq.~\eqref{eq:CMIbound}.

The lesson is that given the channel and the initial and final states, it is still a nontrivial task to check whether the CMI condition Eq.~\eqref{eq:CMIbound} is satisfied ``locally''.

\bibliography{Refs}

\end{document}